\documentclass[conference]{IEEEtran}
\makeatletter
\def\ps@headings{%
\def\@oddhead{\mbox{}\scriptsize\rightmark \hfil \thepage}%
\def\@evenhead{\scriptsize\thepage \hfil \leftmark\mbox{}}%
\def\@oddfoot{}%
\def\@evenfoot{}}
\makeatother
\usepackage{graphicx}
\usepackage{times}
\usepackage{graphics}
\usepackage{makecell}
\usepackage{lipsum}
\usepackage{url}
\usepackage{subfig}
\usepackage{epsfig}
\usepackage{cancel}
\usepackage{caption}
\usepackage{varwidth}
\usepackage{algorithm}
\usepackage{algorithmicx,algpseudocode}

\usepackage[export]{adjustbox}

\usepackage{mwe} 

\newcommand{\CASE}[1]{\STATE \textbf{case} #1\textbf{:} \begin{ALC@g}}
\newcommand{\ENDCASE}{\end{ALC@g}}

\newcommand{\DEFAULT}{\STATE \textbf{default:} \begin{ALC@g}}
\newcommand{\ENDDEFAULT}{\end{ALC@g}}
\newcommand{\DEFAULTLINE}[1]{\STATE \textbf{default:} }

\usepackage{color}
\usepackage{multirow}
\usepackage{multicol}
\usepackage{hhline}
\usepackage{amsthm}
\usepackage{comment}
\usepackage{tabu}
\usepackage{amsmath,amsfonts,amssymb,amsthm,epsfig,epstopdf,url,array}

\theoremstyle{plain}

\theoremstyle{definition}

\theoremstyle{remark}

\IEEEoverridecommandlockouts
\begin{document}

\title{VeriTable: Fast Equivalence Verification of Multiple Large Forwarding Tables}

\author{
Garegin Grigoryan \\ grigorg@clarkson.edu  \\Clarkson University
\and Yaoqing Liu \\ liu@clarkson.edu  \\Clarkson University
\and Michael Leczinsky \\ leczinm@clarkson.edu  \\Clarkson University
\and Jun Li \\ lijun@cs.uoregon.edu \\ University of Oregon
}
\date{}
\maketitle

\pagestyle{empty}

\begin{abstract}

Due to network practices such as traffic engineering and multi-homing, the number of routes---also known as IP prefixes---in the global forwarding tables has been increasing significantly in the last decade and continues growing in a super linear trend. One of the most promising solutions is to use smart Forwarding Information Base (FIB) aggregation algorithms to aggregate the prefixes and convert a large table into a small one. Doing so poses a research question, however, i.e., how can we quickly verify that the original table yields the same forwarding behaviors as the aggregated one? We answer this question in this paper, including addressing the challenges caused by the longest prefix matching (LPM) lookups.
In particular, we propose the VeriTable algorithm that can employ a single tree/trie traversal to quickly check if multiple forwarding tables are forwarding equivalent, as well as if they could result in routing loops or black holes. The VeriTable algorithm significantly outperforms the state-of-the-art work for both IPv4 and IPv6 tables in every aspect, including the total running time, memory access times and memory consumption.

\end{abstract}

\section{Introduction}
\label{sec:intro}

While every router on the Internet has a main Forwarding Information Base (FIB) (i.e. forwarding table) to direct traffic transmission, there are various scenarios where we need to verify if two or more forwarding tables residing in the same or different routers have the same forwarding behaviors. This identical forwarding behavior is also known as forwarding equivalence. The capability to conduct quick, simultaneous equivalence verification on multiple router forwarding tables is vitally important to ensure efficient and effective network operations. We will use two examples to demonstrate the fundamental significance of this research topic.  

First of all, when router vendors develop, test and run their router software and hardware, they must verify that the Forwarding Information Base (FIB) table in the data plane is correctly derived from the Routing Information Base (RIB) table in the control plane. A typical carrier-grade router consists of three primary components: a control engine running various routing protocols, collecting routing information and selecting the best routes to a master forwarding table; many pieces of parallel forwarding engines, called line cards; and a switch fabric linking the control engine and forwarding engines~\cite{Chao:2007:HPS:1202844}. Based on such distributed system design, routers can achieve better scalability and reliability. This also results in at least three copies of the same forwarding table within a single router. One copy is in the control plane, also known as the master forwarding table, which contains the best routes selected by the RIB. Another copy, mirrored from the master forwarding table, resides in the local memory of each line card. The third copy is maintained in each forwarding ASIC chip, which is in charge of fast IP routing lookup and packet forwarding. In theory, the three copies of forwarding tables should have exactly identical forwarding behaviors. However, in reality, this may not always be true. Thus, we are required to use a highly efficient forwarding table verification scheme for debugging and diagnosis purposes. Moreover, routes are frequently updated by neighbors and these changes need to be simultaneously reflected in all three copies of the forwarding table, which makes fast verification between the copies more challenging. For example, Cisco Express Forwarding (CEF) relies on real-time consistency checkers to discover prefix inconsistencies between RIB and FIB (\cite{inconsistency, consistencychecker}), due to the asynchronous nature of the distribution mechanism for both databases.


Second, when Internet Service Providers (ISPs) use FIB aggregation techniques to reduce FIB size on a linecard, they must ensure that the aggregated FIB yields 100\% forwarding equivalence as the original FIB (\cite{draves1999constructing},~\cite{uzmi2011smalta},~\cite{liu2013fifa}). 
The basic idea is that multiple prefixes which share the same next hop or interface can be merged into one. The best routes that are derived from routing decision processes, e.g., BGP decision process, will be aggregated according to the distribution of their next hops before they are pushed to the FIB. The aggregated copy of the routes with a much smaller size will then be downloaded to the FIB. Unlike many other approaches that require either architectural or hardware changes (\cite{mcpherson2009intra},~\cite{saucez2012designing}), FIB aggregation is promising because it is a software solution, local to single routers and does not require coordination between routers in a network or between different networks. This actually leads to an essential question, how can we quickly verify that the different FIB aggregation algorithms yield results which have the same semantical forwarding as the original FIB, particularly in the case where we need to handle many dynamic updates? Therefore, quick simultaneous equivalence verification on multiple forwarding tables is of great importance to verify the correctness of FIB aggregation algorithms' implementation. Although the real-time requirement of equivalence verification is not very high here, it yields great theoretical value to design advanced algorithms to reduce CPU running time.  

More generally, service providers and network operators may want to periodically check if two or more forwarding tables in their network cover the same routing space. Ideally, all forwarding tables in the same domain are supposed to yield the same set of routes to enable reachability between customers and providers. Otherwise, data packets forwarded from one router may be dropped at the next-hop receiving router, also known as ``blackholes". The occurrence of blackhole may stem from multiple reasons, such as misconfigurations, slow network convergence, protocol bugs and so forth. To this end, another essential question is that how we can quickly check if two or more forwarding tables cover the same routing space with consistent routes?


There are at least three challenges to overcome. An efficient algorithm must be able to: 

(1) Verify forwarding equivalence over the entire IP address space, including 32-bit IPv4 and 128-bit IPv6, using the Longest Prefix Matching (LPM) rule in a super-fast manner. LPM rule refers to that the most specific routing entry and the corresponding next hop will be selected when there are multiple matches for the same packet. For example, in Table~\ref{tab:fib1}, one packet destined to $01100011$ (assume 8-bit address space) has two routing matches: $"\_"$ (0/0) with next hop $A$ and $01$ with next hop $B$. However, only the longest prefix $01$ and the next hop $B$ will be used to forward the packet out. When we refer to forwarding equivalence, the next hops, derived from LPM lookups against all participating forwarding tables should be identical for every IP address, thus we need to cover the entire IP address space ($2^{32}$ addresses for IPv4 and $2^{128}$ addresses for IPv6) quickly to check if the condition is satisfied. 


(2) Handle very large forwarding tables with a great number of routing entries. For instance, current IPv4 forwarding table size has been over 700,000 entries~\cite{cidr}. IPv6 forwarding tables are fast growing in a super-linear trend (more than 40,000 entries as of July 2017)~\cite{bgpipv6}. It is estimated that millions of routing entries will be present in the global forwarding tables in the next decade~\cite{bgpprojection}. Can we scale up our verification algorithm to handle large forwarding tables efficiently?

(3) Mutually verify the forwarding equivalence over multiple forwarding tables simultaneously. For example, Table~\ref{tab:fibentry} shows two forwarding tables with different prefixes and next hops, how can we quickly verify whether they yield forwarding equivalence or not under the aforementioned LPM rule? Can we scale out our algorithm to deal with many tables together but still yield good performance?

%
%
%
%
%

\begin{table}[tbp]
	\centering
	\small
	\caption{FIB Table Forwarding Equivalence}
	\captionsetup{justification=centering}
	\label{tab:fibentry}
	\subfloat[FIB table 1]{
		\begin{tabular}{| c | c | }
			\hline
			Prefix  & Next hop   \\ \hline
			- & A    \\ \hline
			000 & B    \\ \hline
			01 & B   \\ \hline
			11 & A    \\ \hline
		    1011 & A    \\ \hline
		\end{tabular}
		\label{tab:fib1}
	}
	\qquad
	\subfloat[FIB table 2]{
		\begin{tabular}{| c | c | }
			\hline
			Prefix  & Next hop   \\ \hline
			- & B    \\ \hline
			001 & A    \\ \hline
			1 & A   \\ \hline
			100 & A   \\ \hline
			
		\end{tabular}
		\label{tab:fib2}
	}
%

	\vspace{-10mm}
	
\end{table}

In this work, we conquered all of the challenges and made the following contributions: (1) We have designed and implemented a new approach to verify \textbf{multiple snapshots of arbitrary routing/forwarding tables} simultaneously through a single PATRICIA Trie~\cite{morrison1968patricia} traversal; (2) It is the first time that we examined the forwarding equivalence over both real and large IPv4 and IPv6 forwarding tables; 
(3) The performance of our algorithm \textit{VeriTable}  significantly outperforms existing work \textit{TaCo} and \textit{Normalization}. For \textit{TaCo}, \textit{VeriTable} is \textbf{2 and 5.6 times faster} in terms of verification time for IPv4 and IPv6, respectively, while it only uses \textbf{36.1\% and 9.3\% of total memory} consumed by \textit{TaCo} in a two-table scenario. For \textit{Normalization} approach, \textit{VeriTable} is \textbf{1.6 and 4.5 times} faster for their total running time for IPv4 and IPv6, respectively; and (4) In a relaxed version of our verification algorithm, we are able to quickly test if multiple forwarding tables cover the same routing space. If not, what are the route leaking points?



This paper has been structured as follows: first we  introduce the state-of-the-art verification algorithm \textit{TaCo} and \textit{Normalization} between two tables in Section~\ref{sec:stateart}; then we present our work \textit{VeriTable} with a compact tree data structure. We depict the design of the algorithm in Section~\ref{sec:ourwork}. We evaluate both IPv4 and IPv6 forwarding tables in terms of running time, number of memory accesses as well as memory usage in the scenarios with two and more tables in Section~\ref{sec:evaluation}. We describe related work in Section~\ref{sec:related}. Finally, we conclude the paper with future work in Section~\ref{sec:conclusion}. 

\section{state of the art}
\label{sec:stateart}
To the best of our knowledge, there are two state-of-the-art algorithms designed for Forwarding Table Equivalence Verification: \textit{TaCo}~\cite{tariq2011taco} and \textit{Normalization}~\cite{retvari2013compressing}.
\subsection{TaCo}
\label{subsec:stateartTaCo}
\begin{figure}[t!]
	\centering
	\captionsetup{position=top}

	\subfloat[Table 1\label{fig:TacoBinaryA}]{
		\includegraphics[width=0.21\textwidth,valign=t]{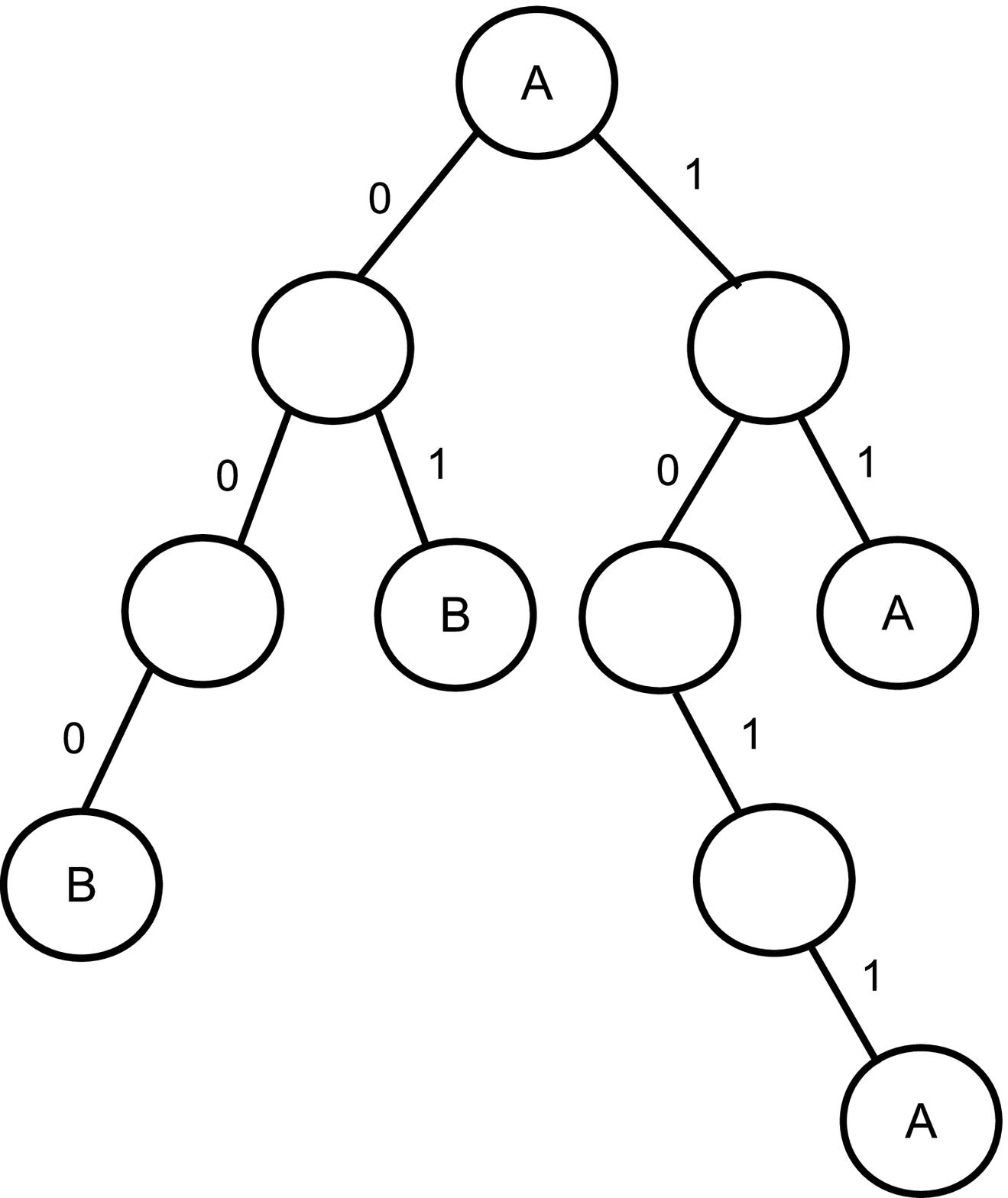}
		}
	\subfloat[Table 2\label{fig:TacoBinaryB}]{
		\includegraphics[width=0.135\textwidth,valign=t]{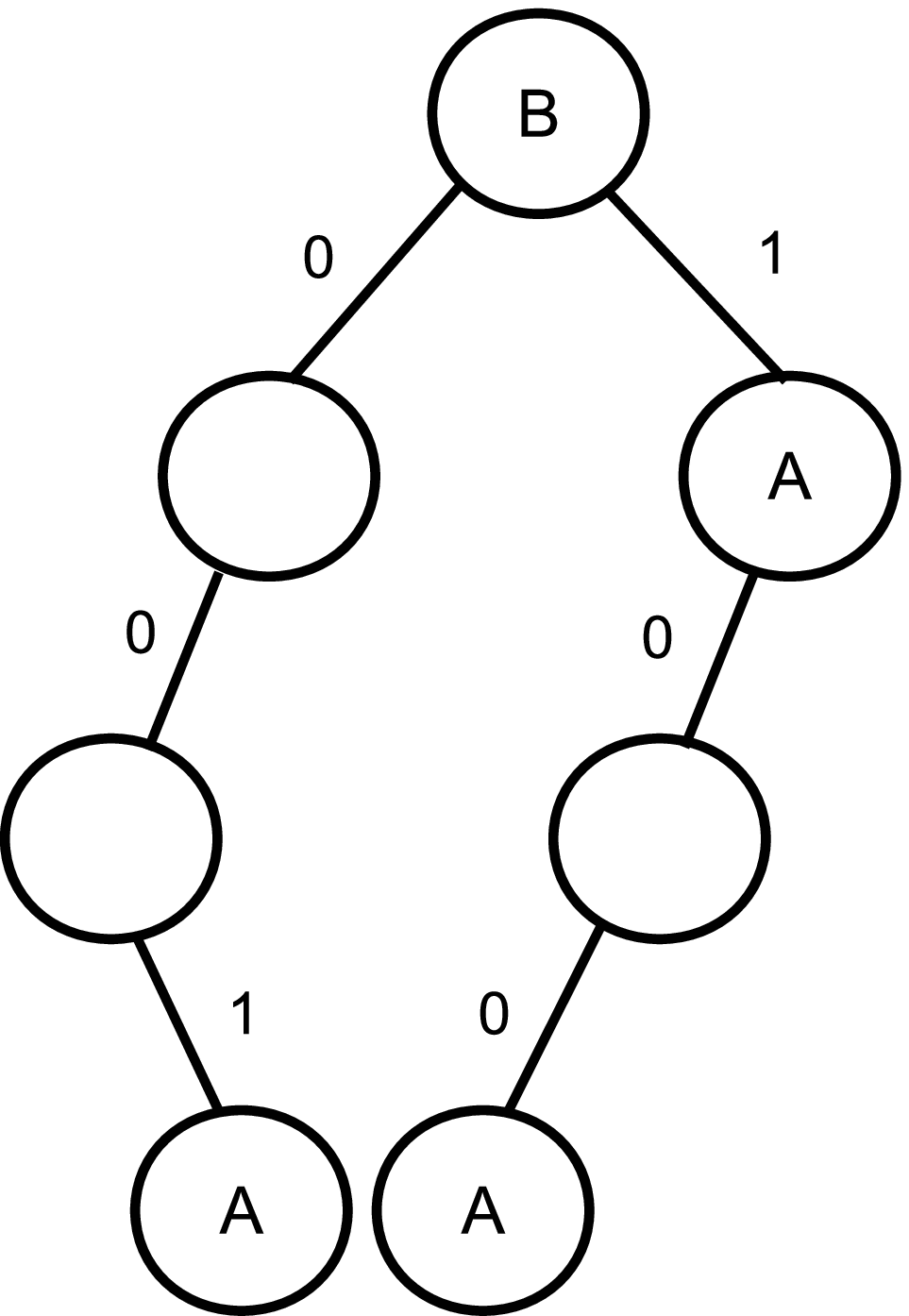}
	}
	
	\caption{Binary Prefix Trees}
	\label{fig:TacoBinary}
	\vspace{-6mm}

\end{figure}

\begin{figure}[t!]
	\centering
	\captionsetup{position=top}
	
	\subfloat[Table 1\label{fig:TacoBinaryNormA}]{
		\includegraphics[width=0.21\textwidth,valign=t]{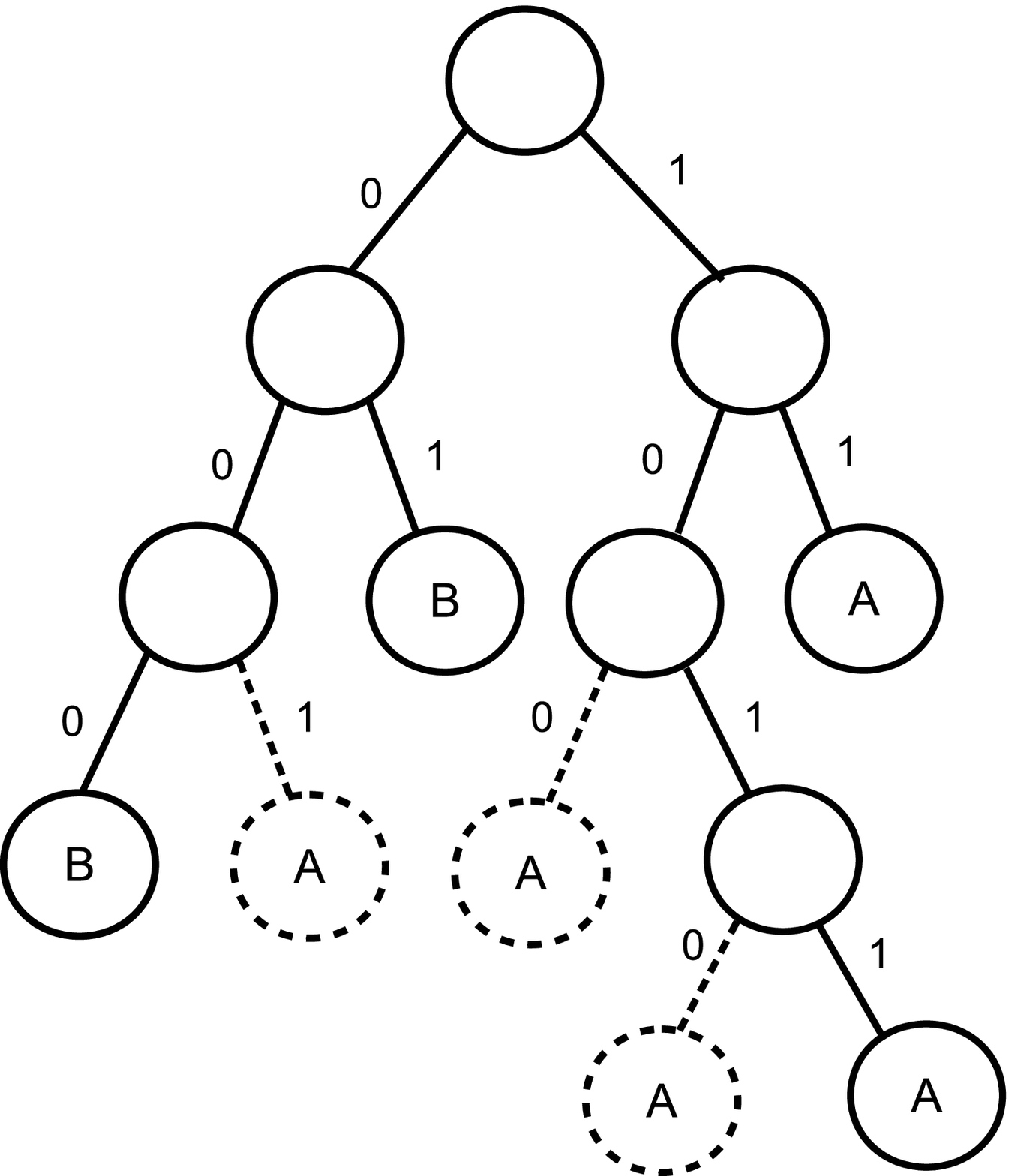}
	}
	\subfloat[Table 2\label{fig:TacoBinaryNormB}]{
		\includegraphics[width=0.18\textwidth,valign=t]{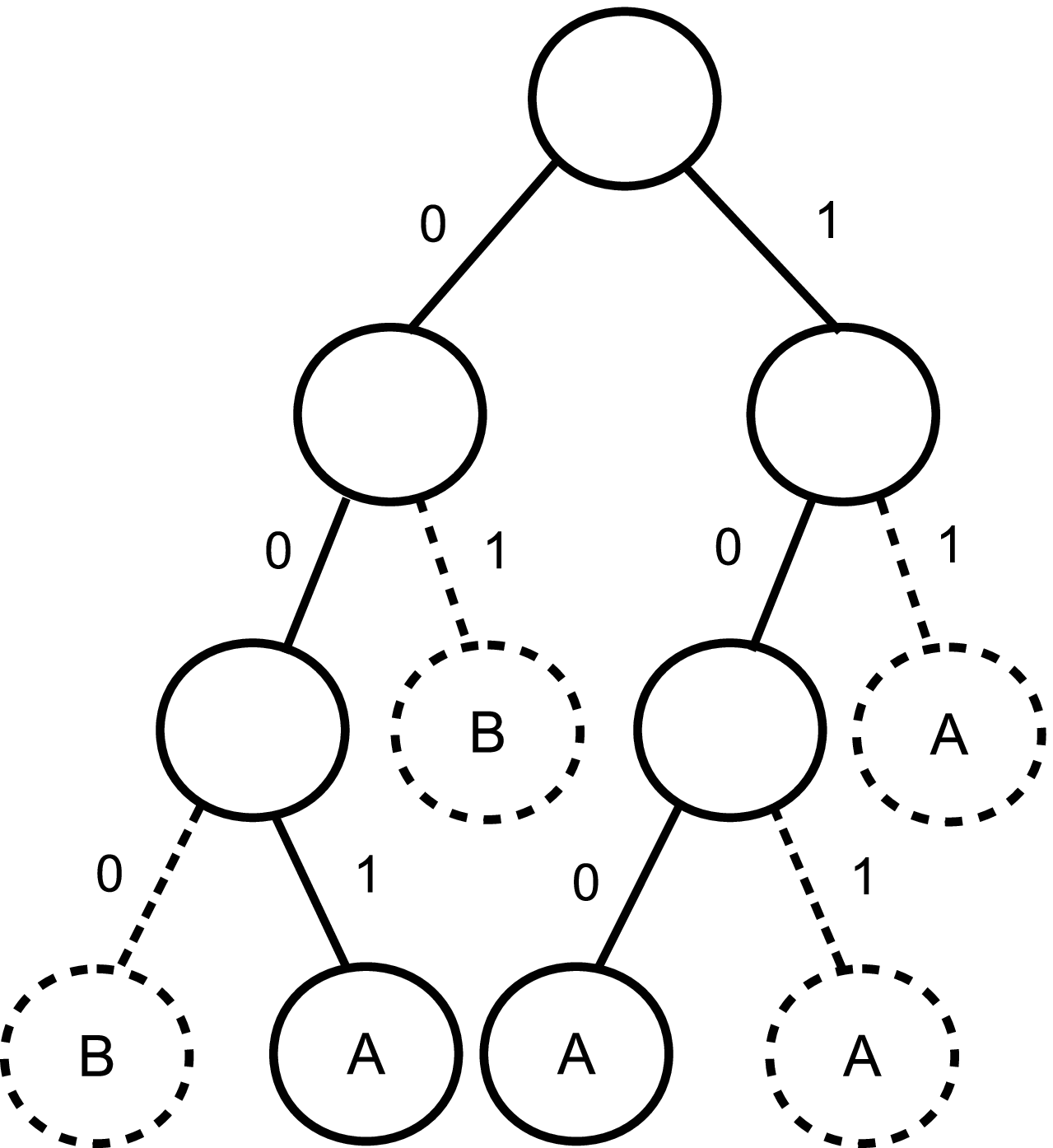}
	}
	
	\caption{Binary Prefix Trees After Leaf Pushing}
	\label{fig:TacoBinaryNorm}
	\vspace{-5mm}
	
\end{figure}

\textit{TaCo} verification algorithm bears the following features: (1) Uses separate Binary Trees (BTs) to store all entries for each forwarding/routing table; (2) Was designed to compare only two tables at once; (3) Has to leverage tree  leaf pushing to obtain semantically equivalent trees; and (4) Needs to perform two types of comparisons: direct comparisons of the next hops for prefixes common
between two tables and comparisons that involve LPM lookups of the IP addresses, extended from the remaining prefixes of each table. More specifically, \textit{TaCo} needs to use four steps to complete equivalence verification for the entire routing space for Tables~\ref{tab:fib1} and~\ref{tab:fib2}: (a) Building a Binary Tree (BT) for each table, as shown in Figure~\ref{fig:TacoBinary}; (b) BT Leaf Pushing, Figure~\ref{fig:TacoBinaryNorm} shows the resultant BTs; (c) Finding common prefixes and their next hops, then making comparisons; and (d) Extending non-common prefixes found in both BTs to IP addresses and making comparisons. Due to space constraint, we skip the detailed process. 


%
%

\begin{figure}[t!]
	\centering
	\captionsetup{position=top}
	
	\subfloat[Table 1\label{fig:NormA}]{
		\includegraphics[width=0.19\textwidth,valign=t]{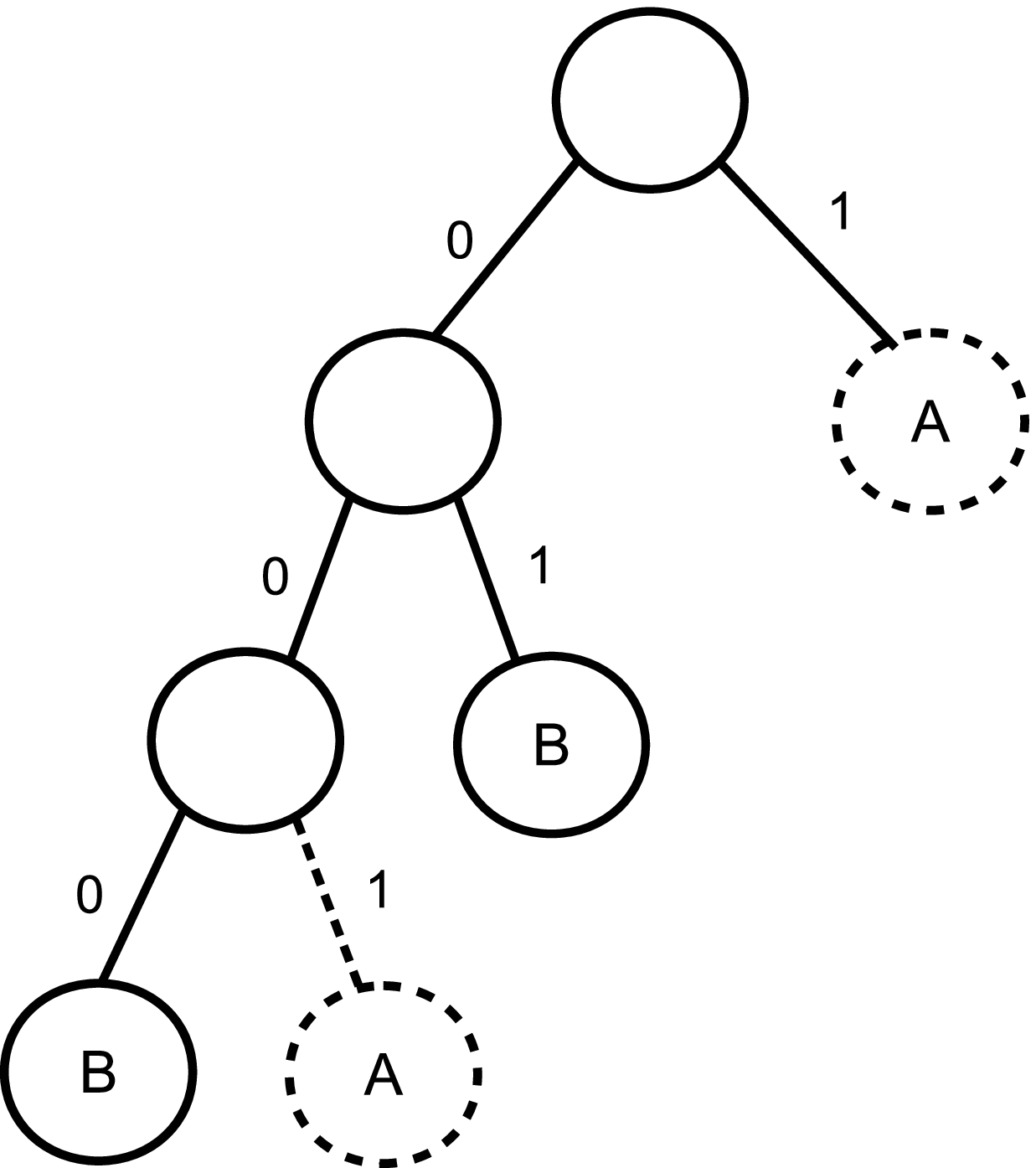}
	}
	\subfloat[Table 2\label{fig:NormB}]{
		\includegraphics[width=0.18\textwidth,valign=t]{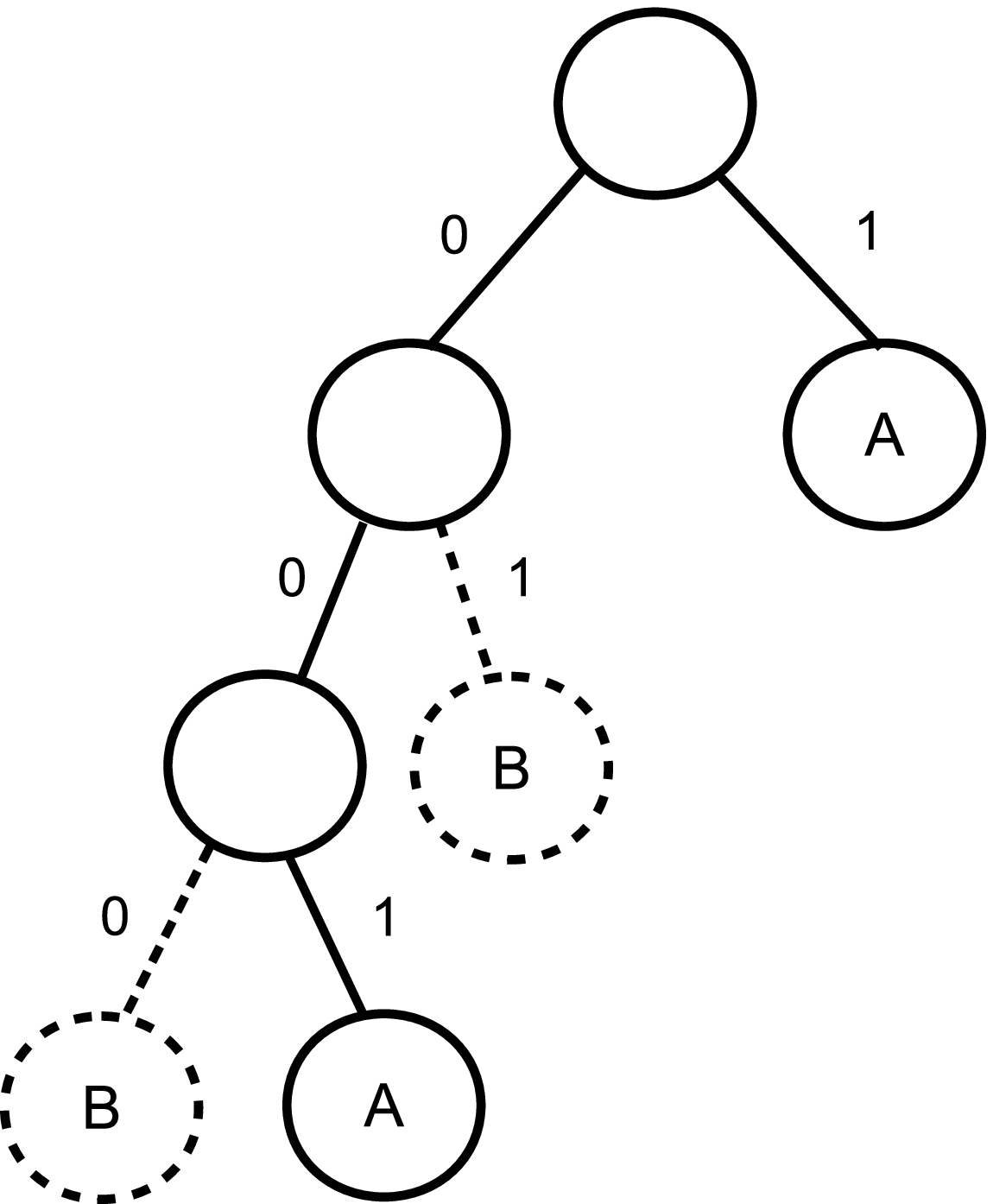}
	}
	
	\caption{Binary Prefix Trees After Normalization}
	\label{fig:normalized}
	\vspace{-10mm}
	
\end{figure}

Finally, when all comparisons end up with the equivalent results, \textit{TaCo} theoretically proves that the two FIB tables yield semantic forwarding equivalence. As a result, \textit{TaCo} undergoes many inefficient operations: ($a$) BT leaf pushing is time and memory consuming; ($b$) To find  common prefixes for direct comparisons \textit{TaCo} must traverse all trees and it is not a trivial task; ($c$) IP address extension and lookups for non-common prefixes are CPU-expensive; and ($d$) To compare $n$ tables and find the entries that cause possible nonequivalence, it may require $(n-1)*n$ times of tree-to-tree comparisons (example: for 3 tables $A$, $B$, $C$ there are 6 comparisons: $A$ vs $B$, $A$ vs $C$, $B$ vs $C$, $B$ vs $A$, $C$ vs $B$, $C$ vs $A$ ). For instance, it may require 90 tree-to-tree combinations to compare 10 tables mutually. On the contrary, our work-\textit{VeriTable} eliminates all these expensive steps and accomplishes the verification over an entire IP routing space through a single tree/trie traversal. We describe it in detail in Section~\ref{sec:ourwork}. 


\subsection{Tree Normalization}
\label{subsec:Unique}
R\'etv\'ari $et$ $al.$ in~\cite{retvari2013compressing} show that a unique form of a BT can be obtained through \textit{Normalization}, a procedure that eliminates brother leaves with identical labels (e. g. next hop values) from a leaf-pushed BT. Indeed, if a recursive substitution is applied to the BTs in Figure~\ref{fig:TacoBinaryNorm}, BT (a) and (b) will be identical. It is possible to prove that the set of tables with the same forwarding behavior have identical normalized BTs. More specifically, \textit{Normalization} verification approach has three steps involved: leaf pushing, tree compression, and side-by-side verification. Leaf pushing refers to formalizing a normal binary tree to a full binary tree, where a node is either a leaf node or a node with two children nodes. The leaf nodes' next hops are derived from their ancestors who have a next hop value originally. Tree compression involves compressing two brother leaf nodes that have the same values to their parent node. This is a recursive process until no brother leaf nodes have the same next hops. The final verification process will traverse every leaf node on both BTs to verify the forwarding equivalence side by side. Figure~\ref{fig:normalized} shows the normalized BTs after the first two steps.
As we can observe, normalizing BTs to a unique form involves several inefficient operations including time-consuming leaf pushing and complicated leaf compression. Section~\ref{sec:evaluation} presents the evaluation results between our algorithm and both \textit{TaCo} and \textit{Normalization} approaches.


 \section{Design of VeriTable}
 \label{sec:ourwork}
 
 In this section, we introduce the main data structures used in our work first, then describe the terms, design steps and the algorithms of \textit{VeriTable}. Along with the presentation, we use an example to show how the verification scheme works. 
 
 \subsection{PATRICIA Trie}
 \label{trie}
 
\begin{figure}[t]
 	\centering
 	\captionsetup{position=top}
 	
 	\subfloat[t][Table 1\label{fig:Table1Patricia}]{
 		\includegraphics[width=0.24\textwidth,valign=t]{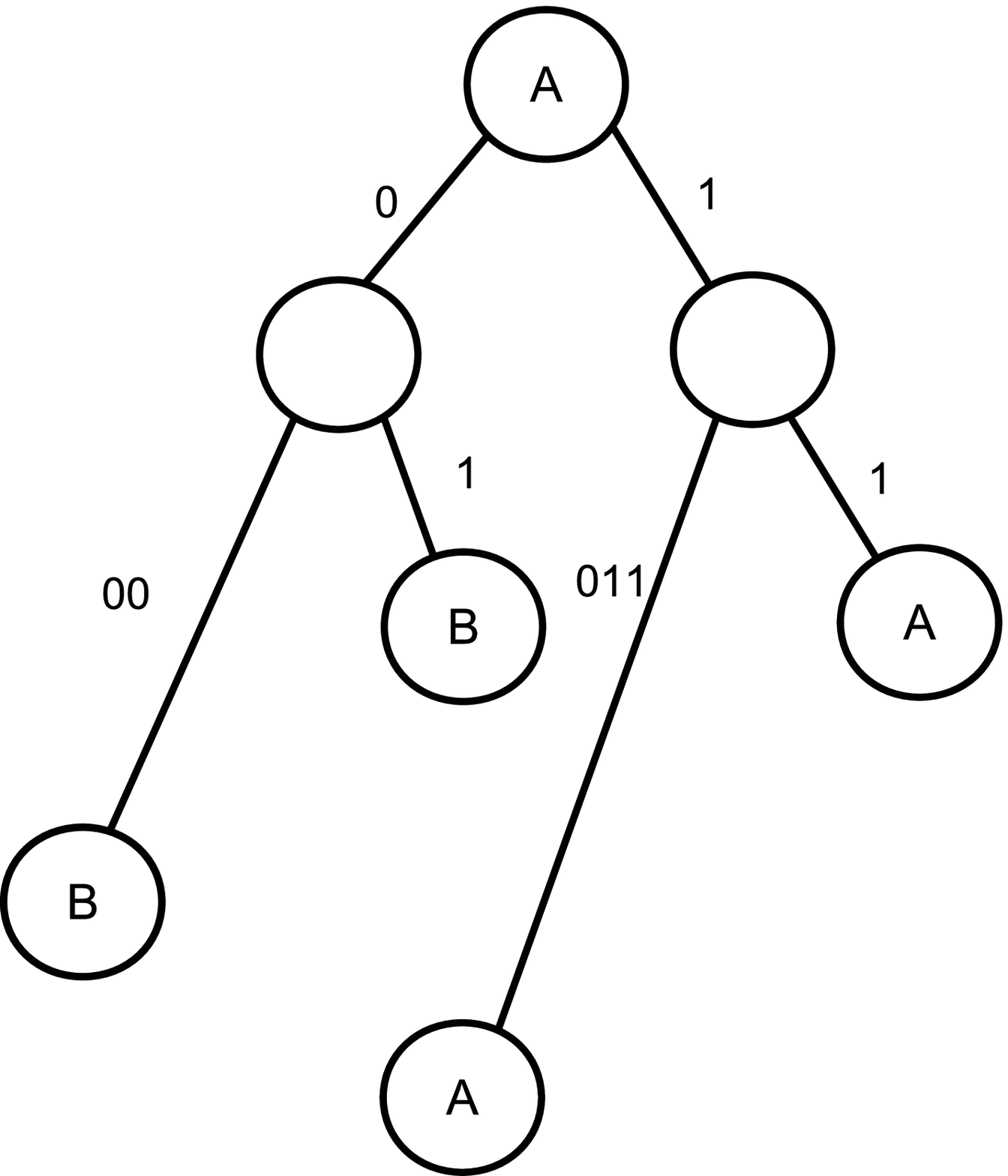}
 	}
 	\subfloat[t][Table 2\label{fig:Table2Patricia}]{
 		\includegraphics[width=0.155\textwidth,valign=t]{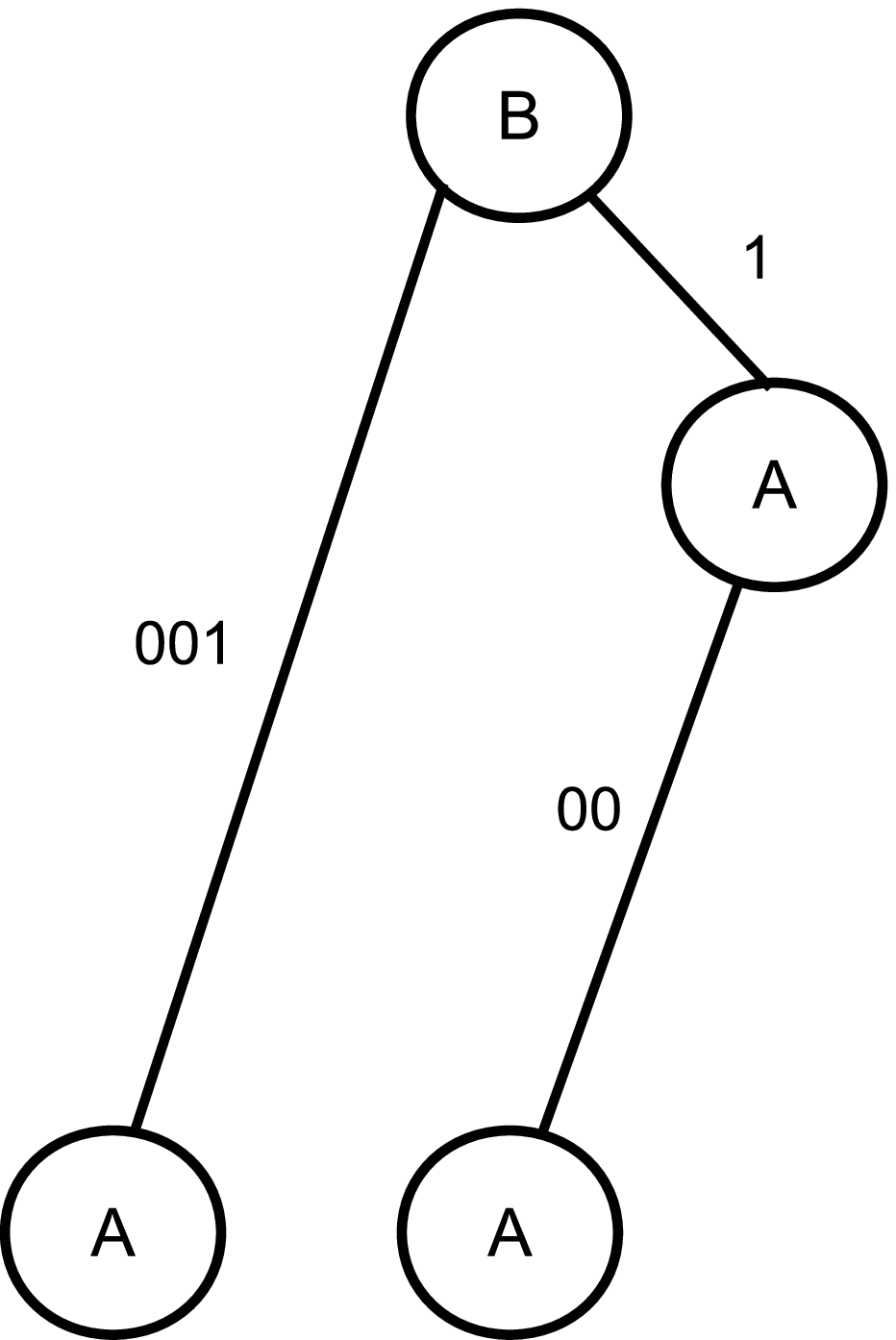}
 		
 	}
 	\vspace{-2mm}\\
 	{\footnotesize Hollow nodes denote \textit{GLUE} nodes helping build the trie structure, other non-hollow nodes are called \textit{REAL} nodes, whose prefixes were from one of the forwarding tables}
 	
 	\caption{PATRICIA Tries (PTs)}
 	\label{fig:patriciatries}
    \vspace{-5mm}
 	
 \end{figure}

 Instead of using a BT to store a forwarding table, we use a PATRICIA (Practical Algorithm to Retrieve Information Coded in Alphanumeric) Trie~\cite{morrison1968patricia} data structure, which is based on a radix tree using a radix of two. PATRICIA Trie (PT) is a compressed binary tree and can be quickly built and  perform fast IP address prefix matching. For instance, Figure~\ref{fig:patriciatries} demonstrates the corresponding PTs for FIB Table~\ref{tab:fib1} and FIB Table~\ref{tab:fib2}. The most distinguished part for a PT is that the length difference between a parent prefix and its child prefix can be equal to and greater than 1. This is different than a BT, where the length difference must be 1. As a result, as shown in the example, PTs only require 7 and 4 nodes, but BTs require 10 and 7 nodes for the two tables, respectively. While the differences for small tables are not significant, however, they are significant for large forwarding tables with hundreds of thousands of entries. An exemplar IPv4 forwarding table with 575,137 entries needs 1,620,965 nodes for a BT, but only needs 1,052,392 nodes for a PT. We have detailed comparison in terms of running time and memory consumption in Evaluation Section~\ref{sec:evaluation}. These features enable a PT to use less space and do faster search, but results in more complicated operations in terms of node creations and deletions, e.g., what if a new node with prefix \textit{100} needs to be added in Figure~\ref{fig:patriciatries}a? It turns out that we have to use an additional glue node to accomplish this task. There are a number of other complex cases, but out of the scope of this work.

  \begin{table}[tbp]
  	\centering
  	\small
  	\caption{Trie Node's Attributes}
  	\vspace{-2mm}
  	\label{tab:trienode}
  	\captionsetup{justification=centering}
  	\begin{tabular}{| l | l | p{4.8cm} |}
  		\hline
  		Name & Data Type & Description   						\\ \hline 	\hline
  		$parent$ & Node Pointer & Points to a node's parent node						\\ \hline
  		$l$ & Node Pointer & Points to a node's left child node	if exists		\\ \hline
  		$r$ & Node Pointer  & Points to a node's right child node if exists		\\ \hline
  		$prefix$& String & Binary string
  		\\ \hline
  		$length$  & Integer & The length of the prefix, 0-32 for IPv4 or 0-128 for IPv6
  		\\ \hline
  		$nexthop$  & Integer Array & Next hops of this prefix in $T_{1}...T_{n}$, size $n$ 	\\ \hline
  		$type$  & Integer & Indicates if a node is a $GLUE$ or $REAL$  	\\ \hline
  		
  	\end{tabular}
  	\vspace{-7mm}
  \end{table}
  
 \subsection{Design}
 \label{design}
 
 Our design consists of two primary tasks: Building and initializing a joint PT, and verifying forwarding equivalence in a post-order traversal over the joint PT.
 
 \subsubsection{Building a Joint PT}
 \label{sec:jointpt}
 
\begin{figure*}[t!]
  	\centering
  	
  	\subfloat[Initial Joint PT\label{fig:initialjointtrie}]{
  		\includegraphics[width=0.25\textwidth]{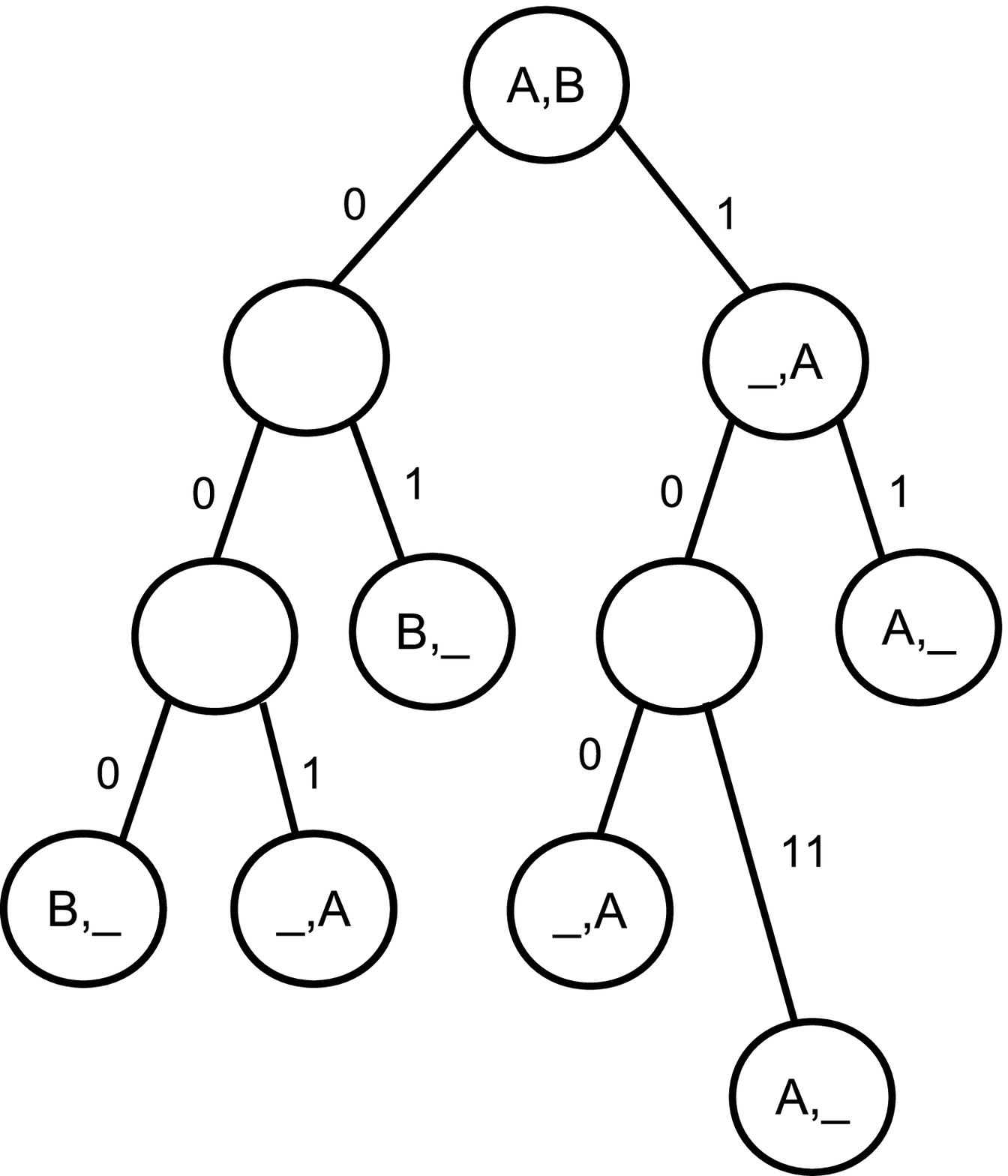}
  	}
  	\subfloat[Joint PT after top-down process \label{fig:topdownjointtrie}]{
  		\includegraphics[width=0.25\textwidth]{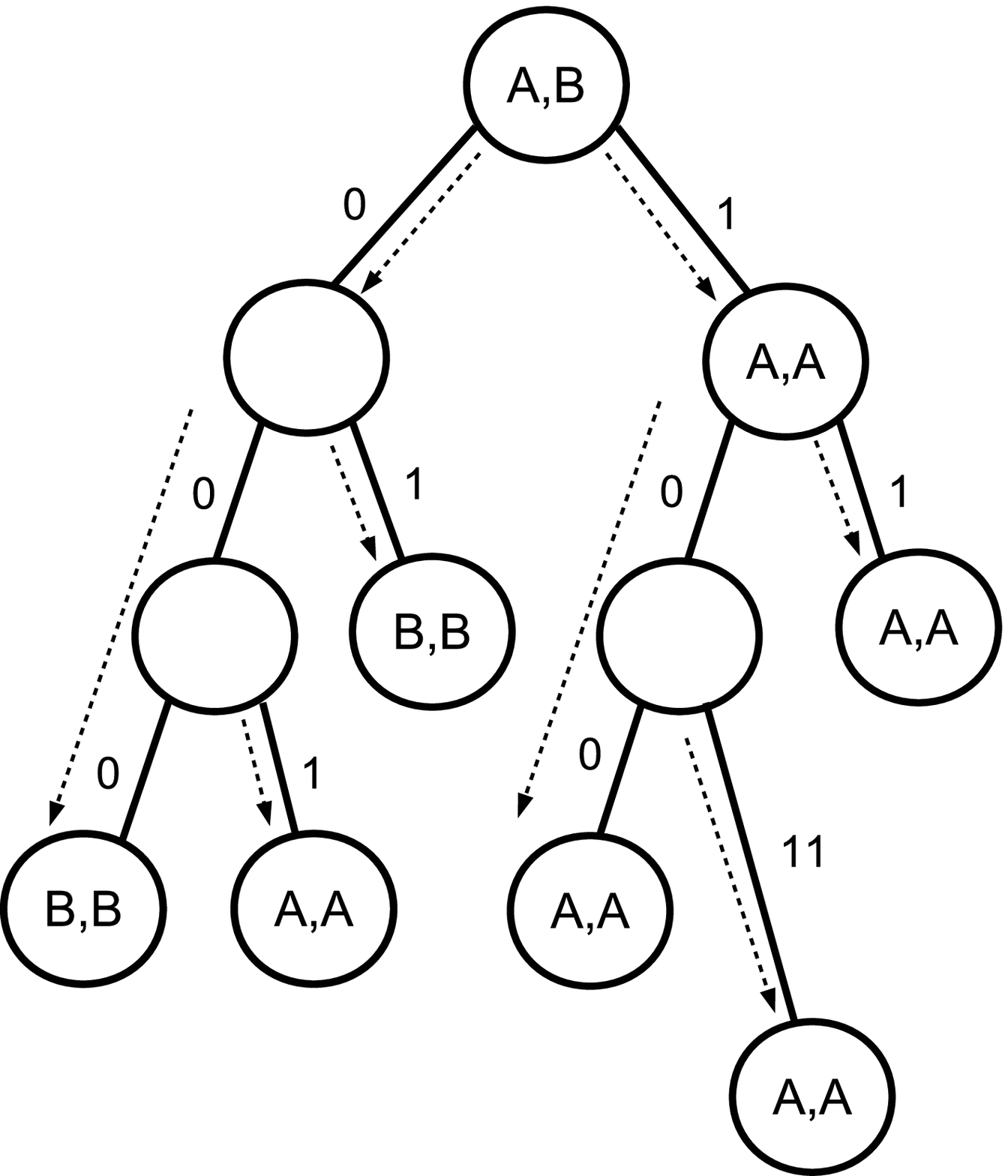}
  	}
  	\subfloat[Joint PT after bottom-up verification \label{fig:bottomupjointtrie}]{
		\includegraphics[width=0.25\textwidth]{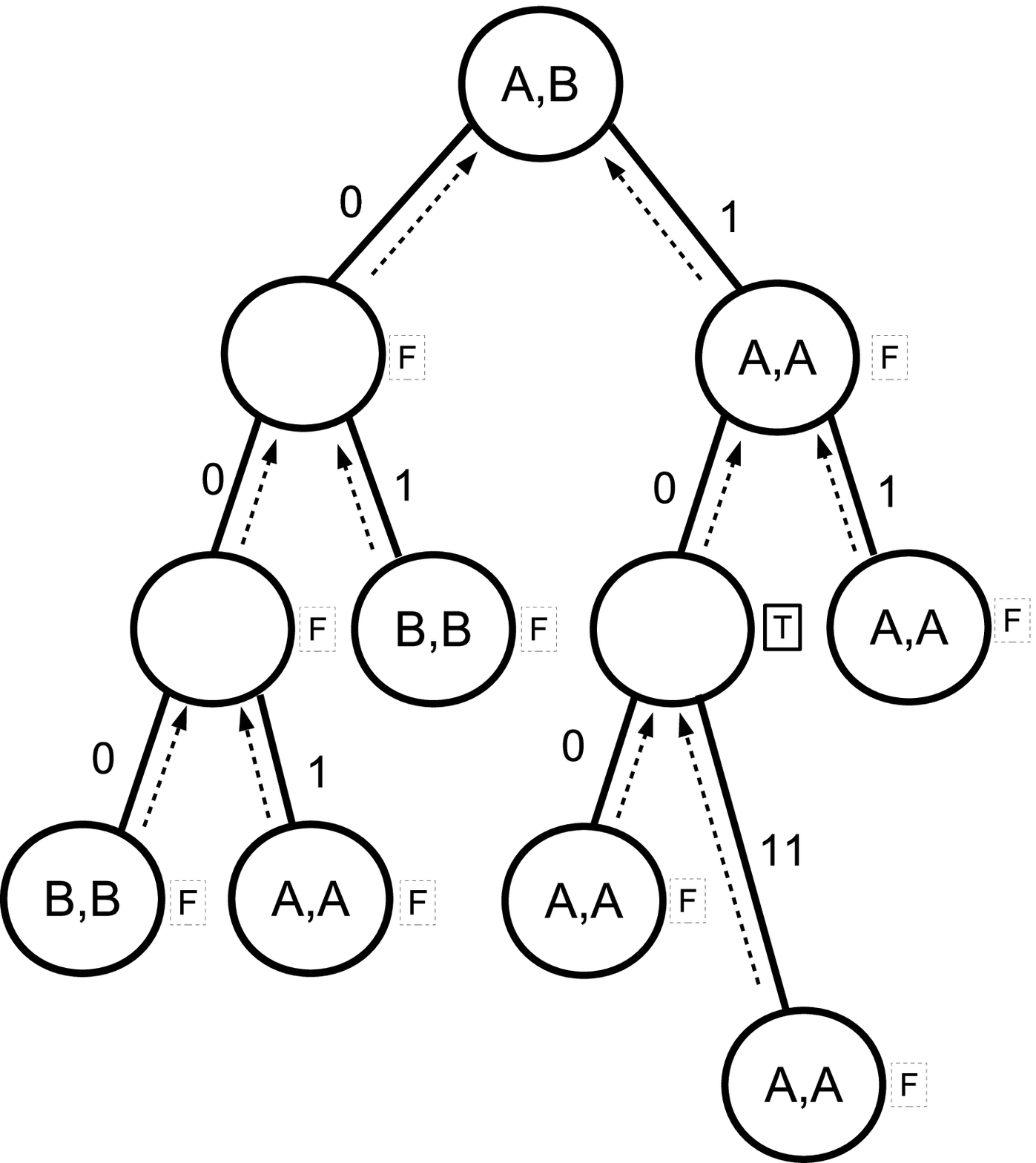}
  		
  	}
  	\vspace{1mm}\\
  	{\footnotesize In Figure $a$, for \textit{REAL} nodes, the $n$th element denotes the next hop value of the corresponding prefix from the $n$th forwarding table. $``\_"$ indicates that no such prefix and next hop exist in the forwarding table. In Figure $b$, after each top-down step, the fields with previous $``\_"$ value will be filled with new next hop values derived from the corresponding \textit{Next Hop} array elements of its nearest \textit{REAL} node. In Figure $c$, $F$ denotes $False$ and $T$ denotes $True$ for the \textit{LEAK} flag. \textit{GLUE} nodes will carry $T$ over to its parent recursively until finding a \textit{REAL} node.}
  	
  	\caption{VeriTable Algorithm}
  	\label{fig:veritable}
  	\vspace{-7mm}
  	
  \end{figure*}

 Building a joint PT for all routing/forwarding tables. Rather than building multiple BTs for each individual table and comparing them in an one-to-one peering manner, as \textit{TaCo} and \textit{Normalization} do, \textbf{we build an accumulated PT using all tables one upon another}. When building the trie, we use a number of fields on each node to help make various decisions. 
 At the beginning, we take the first table as input and initiate all necessary fields to construct a PT accordingly. Afterwards, during the joining process with other tables, the nodes with the same prefixes will be merged. 
 Regarding next hops, we use an integer array to store hops for the same prefix which is located at the same node. The size of the array is the same as the number of tables for comparison. The next hops cannot be merged because they may be different for the same prefix in different tables and also will be used for comparisons, thus they will be placed at the corresponding $n$th element in the array starting from 0, where $n$ is the index number of the input FIB table (we assume only one next hop for each distinct prefix in an FIB table in this work). For instance, the next hop $A$ of prefix \textit{001} in FIB Table $2$ will be assigned as the second element in the \textit{Next Hop} Array on the node with prefix \textit{001}. If there is no next hop for a prefix in a particular table, the element value in the array will be initialized as $``\_"$ by default, or called \textit{``empty"} next hop (we used ``-1" in our implementation). If there is at least one $``non$-$empty"$ next hop in the array, we mark the \textit{Node Type} value as \textit{REAL}, indicating this node contains a real prefix and next hop derived from one of the tables. Otherwise, we call it a \textit{GLUE} node. Algorithm~\ref{alg:patriciabuild} elaborates the detailed workflow to build a joint PT for multiple tables. Table~\ref{tab:trienode} describes a trie node's attributes in our data structure. Figure~\ref{fig:initialjointtrie} shows the resultant joint PT for FIB Table~\ref{tab:fib1} and~\ref{tab:fib2}. 
 

  \algnewcommand\algorithmicforeach{\textbf{for each}}
  \algdef{S}[FOR]{ForEach}[1]{\algorithmicforeach\ #1\ \algorithmicdo}
  
  \begin{algorithm}
  	\caption{Building a Joint PT $T$}\label{alg:patriciabuild}
  	\begin{algorithmic}[1]
  		\Procedure{$BuildJointPT$}{$T_{1..n}$}
  		\State Initialize a PT $T$ with its head node 
  		\State Add prefix $0/0$ on its head node. 
  		\State Set default next hop values in the \textit{Next Hops} array.
  		\ForEach {table $T_{i} \in T_{1..n}$}
  		\ForEach {entry $e$ in $T_{i}$}
  		\State Find a node $n$ in $T$ such as $n.prefix$ is a \par
  		\hskip\algorithmicindent \hskip\algorithmicindent longest match for $e.prefix$ in $T$
  		\If {$n.prefix = e.prefix$}
  		\State $n.nexthop_{i} \gets e.nexthop$
  		\State $n.type \gets REAL$
  		\Else 
  		\State Generate new node $n'$
  		\State $n'.prefix \gets e.prefix$
  		\State $n'.nexthop_{i} \gets e.nexthop$
  		\State $n'.type \gets REAL$
  		\State Assume $n$ has a child $n_{c}$ 
  		\If {the overlapping portion of $n_{c}$  and $n'$ is longer than $n.length$ but shorter than $n'.length$ bits}  
  		\State Generate a glue node $g$
  		\State $n'.parent \gets g$
  		\State $n_{c}.parent \gets g$
  		\State $g.parent \gets n$
  		\State $g.type \gets GLUE$
  		\State Set $g$ as a child of $n$
  		\State Set $n'$ and $n_{c}$ as children of $g$ 
  		\Else
  		\State $n'.parent \gets n$
  		\State $n_{c}.parent \gets n'$
  		\State Set $n_{c}$ as a child of $n'$ 
  		\State Set $n'$ as a child of $n$ 
  		
  		\EndIf
  		
  		\EndIf
  		\EndFor
  		\EndFor
  		\EndProcedure
  		
  	\end{algorithmic}
  \vspace{-1mm}
  \end{algorithm}

%
%
%
 
 There are a few advantages for the design of a joint PT: ($a$) Many common prefixes among different tables will share the same trie node and prefix, which can \textbf{considerably reduce memory consumption and computational time for new node creations}; ($b$) Common prefixes and uncommon prefixes will be automatically gathered and identified in one single tree after the combination; and ($c$) The design will greatly speed up subsequent comparisons of next hops between multiple tables without traversing multiple tries.

 \subsubsection{Post-Order Equivalence Verification}
 \label{sec:verification} 
 After building the joint PT and initializing all necessary fields, we start the verification process, which only needs one post-order PT traversal and includes two steps to accomplish the forwarding equivalence verification as follows:  
 
%
  
 (A) Top-down inheriting next hops. First, we follow a simple but very important rule: \textbf{\textit{According to the LPM rule, the real next hop value for a prefix that has an ``empty" next hop on the joint PT should be inherited from its closest \textit{REAL} ancestor, whose next hop exists and is ``non-empty"}}. For example, to search the LPM matching next hop for prefix \textit{000} in the second table using Figure~\ref{fig:initialjointtrie}, the next hop value should return $B$, which was derived from the second next hop $B$ of its nearest \textit{REAL} ancestor -- the root node. The top-down process will help each specific prefix on a \textit{REAL} node in the joint PT to inherit a next hop from its closest \textit{REAL} ancestor if the prefix contains an ``empty" next hop. More specifically, when moving down, we compare the \textit{Next Hops} array in the \textit{REAL} ancestor node with the array in the \textit{REAL} child node. If there are elements in the child array with ``empty" next hops, then our algorithm fills them out with the same values as the parent. If there are ``non-empty" next hops present in the child node, then we keep them. Note that all \textit{GLUE} nodes (hollow nodes in Figure~\ref{fig:initialjointtrie}) will be skipped during this process, because they are merely ancillary nodes helping build up the trie structure and do not carry any next hop information. After this step, every \textit{REAL} node will have a new \textit{Next Hops} array without any \textit{``empty"} next hops. \textbf{The instantiated next hops will facilitate the verification process without additional retrievals of next hops from their distant ancestors.}  Figure~\ref{fig:topdownjointtrie} shows the results after the top-down step. If there is not a default route 0/0 in the original forwarding tables, we make up one with next hop value 0 and node type \textit{REAL} for calculation convenience.
 
  
 (B) Bottom-up verify LPM next hops. In fact, this process is interwoven with the top-down process in our recursive post-order verification program. While the program moves downward, the top-down operations will be executed. While it moves upward, a series of operations will be conducted as follows. First of all, a leaf node at the bottom may be encountered, where the \textit{Next Hops} array will be checked linearly, element by element. If there are any discrepancies, then we can immediately conclude that the forwarding tables yield different forwarding behaviors, because the LPM prefixes end up with different next hops. In other words, they are not forwarding equivalent. If all next hops share the same value, we move upward to its \textbf{directly connected} parent node, where we check the prefix length difference from the recently visited child node. 
 
 Since we use a PT as our data structure, two cases may occur: $d=1$ and $d>1$, where $d$ denotes the length difference between the parent node and the child node. 
 The first case $d=1$ for all children nodes implies that the parent node has no extra routing space to cover between itself and the children nodes. On the contrary, the second case $d>1$ indicates the parent node covers more routing space than that of all children nodes. 
 If $d>1$ happens at any time, we will set a \textit{LEAK} flag variable at the parent node to indicate that all of the children nodes are not able to cover the same routing space as the parent, which will lead to ``leaking" certain routing space to check for verification. Therefore, in this case, the parent itself needs to be checked by the LPM rule to make sure the ``leaking" routing space is checked as well. If there is no child for a given parent, we consider it as $d>1$. As long as there is one \textit{LEAK} flag initiated at a parent node, the flag will be carried over up to the nearest \textit{REAL} node, which can be the parent node itself or a further ancestor. The verification process of forwarding equivalence will be conducted on the \textit{Next Hops} array of this \textit{REAL} node. Once the process passes over a \textit{REAL} node, the flag will be cleared so that the ``leaking" routing space will not be double checked. Intuitively, we check the forwarding equivalence over the routing space covered by leaf nodes first, then over the remaining ``leaking" routing space covered by internal \textit{REAL} nodes. Figure~\ref{fig:bottomupjointtrie} demonstrates the bottom-up \textit{LEAK} flag setting and carried-over process. For example, $d=2$ between parent \textit{10} and its child \textit{1011}, so the \textit{LEAK} flag on node \textit{10} will be set to $True$ first. Since node \textit{10} is a \textit{GLUE} node, the \textit{LEAK} flag will be carried over to its nearest \textit{REAL} ancestor node \textit{1} with the \textit{Next Hops} array \textit{(A,A)}, where the $leaking$ routing space will be checked and accordingly the \textit{LEAK} flag will be cleared to $False$ to avoid future duplicate checks.

%
%
%
%
%
 
 In Algorithm~\ref{alg:veritable}, we show the pseudocode of our recursive function \textit{VeriTable} that quickly verifies whether the multiple forwarding tables are equivalent or not. If not, the corresponding LPM prefixes and next hops will be printed out. The algorithm consists of both a top-down next hop inheritance process and a bottom-up LPM matching and verification process. We have mathematically proved the correctness of the algorithm but do not present the proof here due to limited space.


\section{Evaluation}
\label{sec:evaluation}

All experiments are run on a machine with Intel Xeon Processor E5-2603 v3 1.60GHz and 64GB memory. Datasets are provided by the RouteViews project of the University of Oregon (Eugene, Oregon USA)~\cite{routeviews}. We collected 12 IPv4 RIBs and 12 IPv6 RIBs on the first day of each month in 2016, and used AS numbers as next hops to convert them into 24 routing/forwarding tables. By the end of 2016, there were about 633K IPv4 routes and 35K IPv6 routes in the global forwarding tables. We then applied an optimal FIB aggregation algorithm to these tables to obtain the aggregated forwarding tables. 
IPv4 yields a better aggregation ratio ({about 25}\%) than IPv6 ({about 60}\%), because IPv4 has a larger number of prefixes. The original and aggregated tables were semantically equivalent and used to evaluate the performance of our proposed \textit{VeriTable} vs the state-of-the-art \textit{TaCo} and \textit{Normalization} (see description in Section~\ref{sec:stateart}) verification algorithms in a two-table scenario. We use the following metrics for the evaluations: tree/trie building time, verification time,  number of node accesses, and memory consumption.


\setlength{\textfloatsep}{8.5pt}
\begin{algorithm}
	\caption{Forwarding Equivalence Verification.
		The initial value of $ancestor$ is $NULL$, and the initial value of $node$ is $T\rightarrow root$. For simplicity, we assume the root node is $REAL$.}\label{alg:veritable}
	\begin{algorithmic}[1]
		
		\Procedure{\textit{VeriTable}}{$ancestor$, $node$}
		
		\If{$ node.type = REAL$}
		\State $ancestor$ = $node$\Comment The closest ancestor node for a REAL node is the node istelf
		\EndIf
		
		\State $l \leftarrow node.l$
		\State $r \leftarrow node.r$
		
		\If{$l \not=NULL$}
		\If{$l.type = REAL$}
		\State $InheritNextHops(ancestor, l)$\Comment A REAL child node inherits next hops from the closest REAL ancestor to initialize ``empty" next hops
		\EndIf
		\State $LeftFlag \leftarrow VeriTable(ancestor, l)$\Comment LeftFlag and RightFlag signify the existing leaks at the branches
		\EndIf
		
		\If{$ r\not=NULL$}
		\If{$r.type = REAL$}
		\State $InheritNextHops(ancestor, r)$
		\EndIf
		\State $RightFlag \leftarrow VeriTable(ancestor, r)$
		\EndIf
		
		\If{$l = NULL\land r = NULL$}
		\State {$CompareNextHops(node)$}\Comment The leaf nodes' next hops are always compared; a verified node always returns the false LeakFlag.
		\State $LeakFlag \leftarrow False$
		\State \Return $LeakFlag$
		\EndIf
		
		\If{$ l \not=NULL \land l.length - node.length > 1 $}
		\State $LeakFlag \leftarrow True$
		
		\ElsIf{$ r \not=NULL \land r.length - node.length > 1 $}
		\State $LeakFlag \leftarrow True$
		
		\ElsIf{$l = NULL \lor r = NULL$}
		\State $LeakFlag \leftarrow True$
		
		\ElsIf{$LeftFlag = True \lor RightFlag = True$}
		\State $LeakFlag \leftarrow True$
		\EndIf
		
		\If{$LeakFlag = True \land node.type = REAL$}
		\State {$CompareNextHops(node)$}
		\State $LeakFlag \leftarrow False$
		\EndIf
		
		\Return $LeakFlag$
		
		\EndProcedure

	\end{algorithmic}
\vspace{-1.5mm}
\end{algorithm}
\subsection{Tree/Trie Building Time}
\label{sec:buildingtime}

\begin{figure}[t]
		\captionsetup{justification=centering, width=0.49\linewidth}
		\subfloat[IPv4 Building Time\label{fig:ipv4buildtime}]
		\centering
		\includegraphics[width=0.49\linewidth]{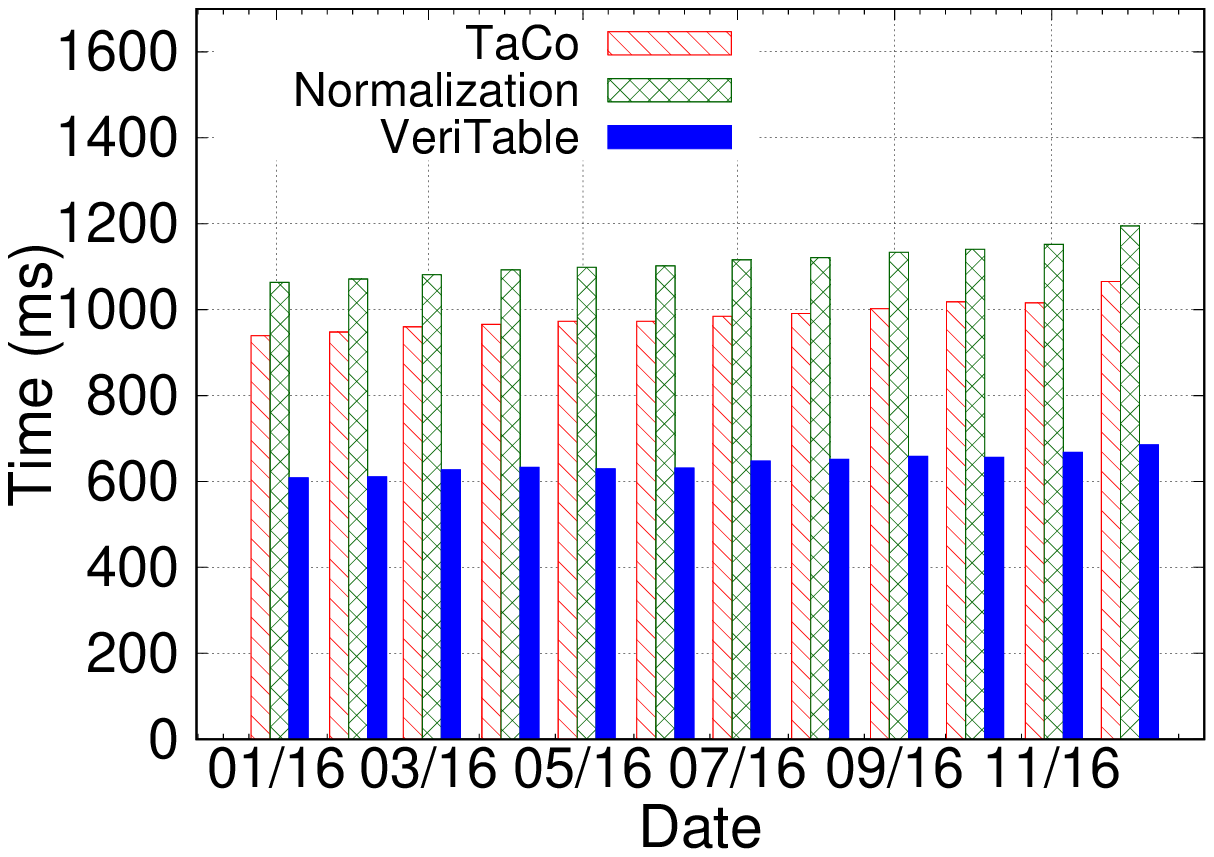}
		\subfloat[IPv6 Building Time\label{fig:ipv6buildtime}]
		\centering
		\includegraphics[width=0.49\linewidth]{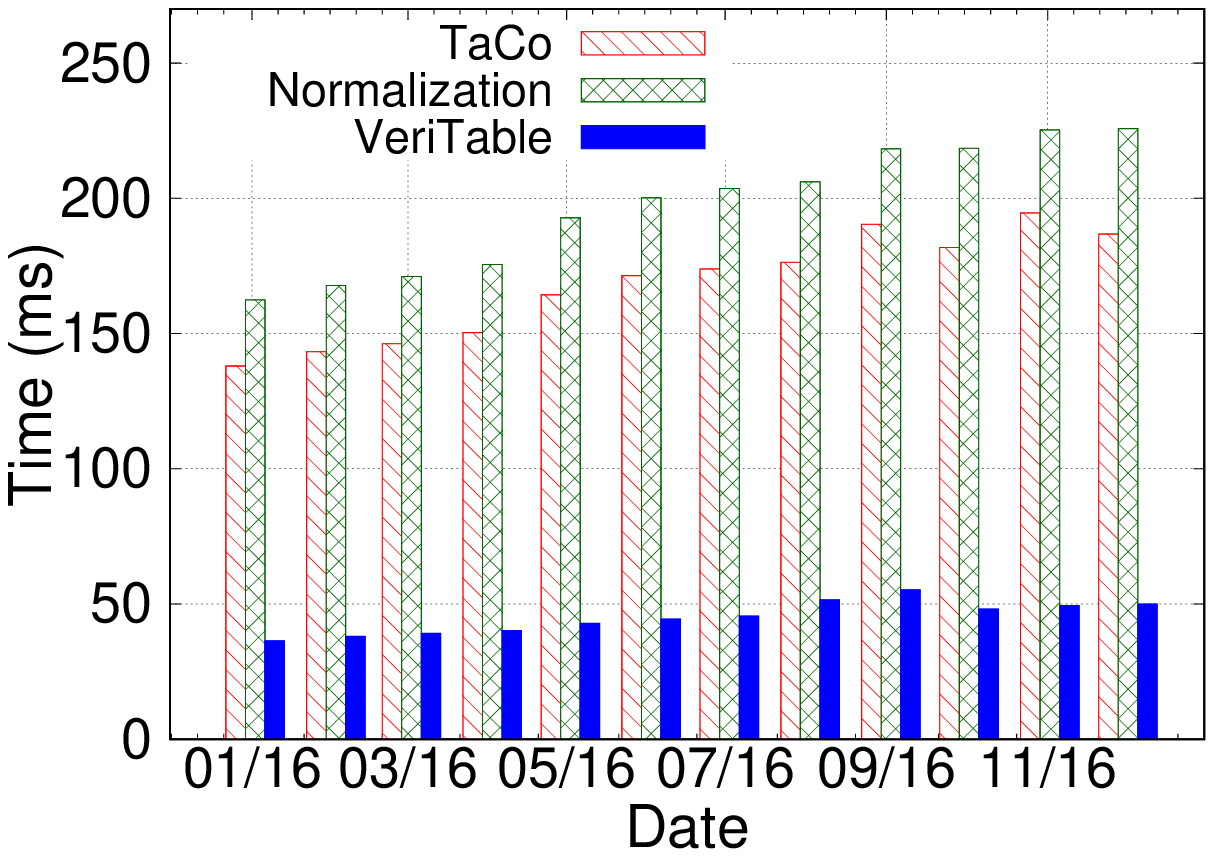}
	\vspace{-3mm}	
	\captionsetup{justification=centering, width=\linewidth}
	\caption{IPv4 and IPv6 Tree/Trie Building Time}
	\label{fig:buildtime}
	\vspace{-4mm}
\end{figure}

\textit{TaCo}, \textit{Normalization} and \textit{VeriTable} need to build their data structures using forwarding table entries before the verification process. \textit{TaCo} and \textit{Normalization} need to build two separate BTs while \textit{VeriTable} only needs to build one joint PT. Figure~\ref{fig:buildtime} shows the building time for both IPv4 and IPv6. Our algorithm \textit{VeriTable} outperforms \textit{TaCo} and \textit{Normalization} in both cases. In Figure~\ref{fig:ipv4buildtime} for IPv4, \textit{TaCo} uses minimum 939.38$ms$ and maximum 1065.41$ms$ with an average 986.27$ms$ to build two BTs. For \textit{Normalization}, it is 1063.42$ms$, 1194.95$ms$ and 1113.96$ms$ respectively. Our \textit{VeriTable} uses minimum 608.44$ms$ and maximum 685.02$ms$ with an average 642.27$ms$ to build a joint PT. \textit{VeriTable} only uses 65.11\% of the building time of \textit{TaCo} and  57.65\% of the building time of \textit{Normalization} for IPv4 tables. In the scenario of IPv6 in Figure~\ref{fig:ipv6buildtime}, \textit{TaCo} uses minimum 137.94$ms$ and maximum 186.73$ms$ with an average 168.10$ms$ to build two BTs; for \textit{Normalization} these numbers are 162.40$ms$, 225.75$ms$ and 197.25$ms$. Our \textit{VeriTable} uses minimum 36.39$ms$ and maximum 49.99$ms$ with an average 45.06$ms$ to build a joint PT. \textit{VeriTable} only uses 26.78\% and 22.84\% of the building time of \textit{TaCo} and \textit{Normalization} respectively for IPv6 tables. Although IPv6 has much larger address space than IPv4, \textit{VeriTable} yields much less building time under IPv6 than that of IPv4, which we attribute to the small FIB size and the usage of a compact data structure -- a joint PT in our algorithm. Note the slower \textit{Normalization} building time due to the operation of tree compression performed by that algorithm.

%
%

\subsection{Verification Time}
\label{sec:verificationtime}

\begin{figure}[t]
		\captionsetup{justification=centering, width=0.49\linewidth}
		\subfloat[IPv4 Verification Time\label{fig:ipv4veritime}]
		\centering
		\includegraphics[width=0.49\linewidth]{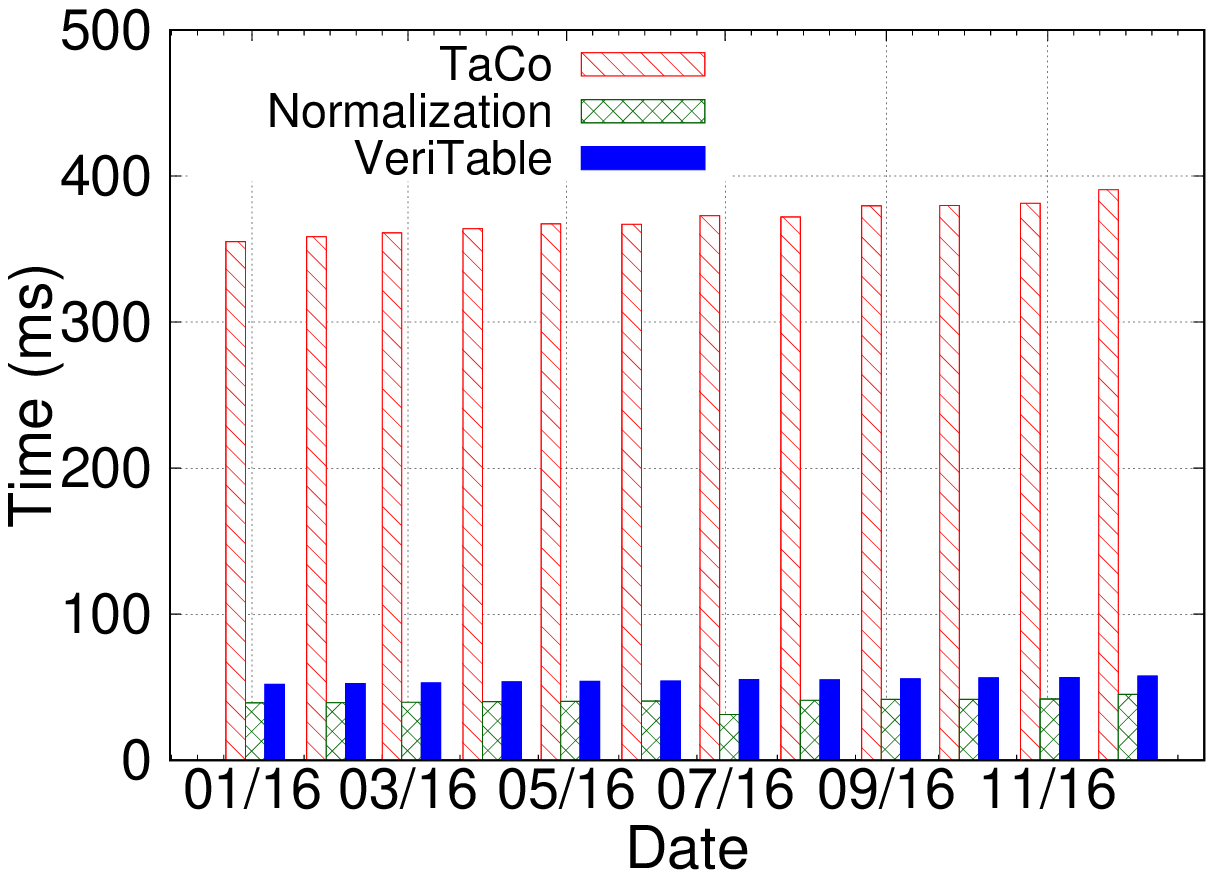}
		\subfloat[IPv6 Verification Time\label{fig:ipv6veritime}]
		\centering
		\includegraphics[width=0.49\linewidth]{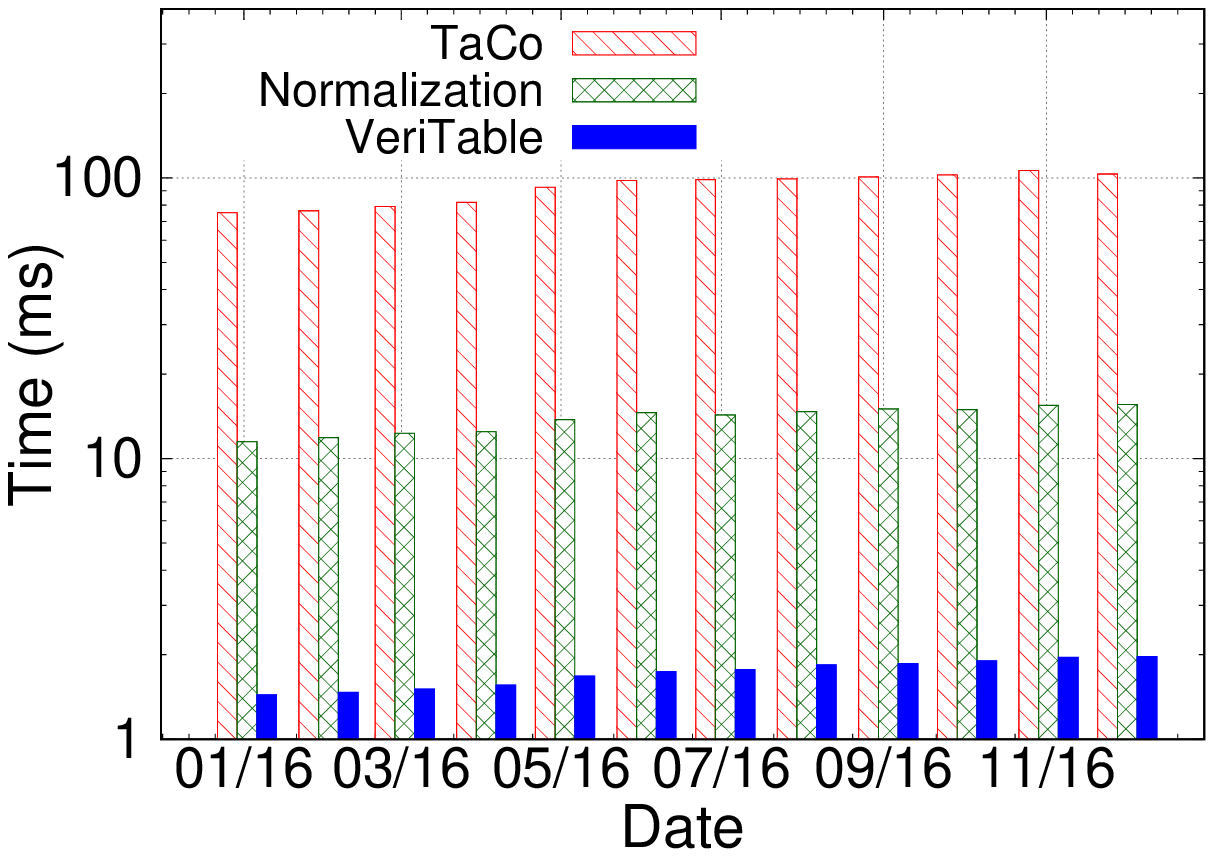}	
	\vspace{-3mm}	
	\captionsetup{justification=centering, width=\linewidth}
	\caption{IPv4 and IPv6 Verification Time}
	\label{fig:veritime}
	\vspace{-4mm}
\end{figure}

A valid verification algorithm needs to cover the whole routing space ($2^{32}$ IP addresses for IPv4 and $2^{128}$ IP addresses for IPv6) to check if two tables bear the same forwarding behaviors. The verification time to go through this process is one of the most important metrics that reflects whether the algorithm runs efficiently or not. Figure~\ref{fig:veritime} shows the running time of \textit{TaCo}, \textit{Normalization} and \textit{VeriTable} for both IPv4 and IPv6, respectively. Our \textit{VeriTable} significantly outperforms \textit{TaCo} in both cases. \textit{TaCo} takes minimum 355.06$ms$ and maximum 390.60$ms$ with an average 370.73$ms$ to finish the whole verification process. Our \textit{VeriTable} takes minimum 51.91$ms$ and maximum 57.48$ms$ with an average 54.63$ms$ to verify the entire IPv4 routing space. \textit{VeriTable} only takes 14.73\% of the verification time of \textit{TaCo} for verification over two IPv4 tables. \textbf{Taking building time into consideration, \textit{VeriTable} is about 2 times faster than \textit{TaCo} for IPv4 verification (1356$ms$ VS 696$ms$)}. 
\textit{Normalization} verification time for IPv4 tables is slightly faster than that of \textit{VeriTable} (which is not the case for IPv6 tables). This is achieved due to the compression that shrinks the size of the BTs for verification process. However, \textit{Normalization} has much longer building time than \textit{VeriTable}.  \textbf{Overall, considering both building and verification time, \textit{VeriTable} is faster than \textit{Normalization} by 40\% (696.90$ms$ VS 1154.08$ms$) for IPv4 verification}. 

Figure~\ref{fig:ipv6veritime} shows the IPv6 scenario (note the Y-axis is a log scale). \textit{TaCo} takes minimum 75.17$ms$ and maximum 103.18$ms$ with an average 92.79$ms$ to finish the whole verification process. For \textit{Normalization}  it is 11.47$ms$, 15.58$ms$, 13.87$ms$ respectively. Our \textit{VeriTable} takes minimum 1.44$ms$ and maximum 1.97$ms$ with an average 1.75$ms$ to verify the entire IPv6 routing space. \textit{VeriTable} only takes 1.8\% and 12.6\% of the verification time of \textit{TaCo} and \textit{Normalization} respectively  for verification over two IPv6 tables. \textbf{Considering both building and verification time, \textit{VeriTable} is 5.6 times faster than \textit{TaCo} (261$ms$ VS 47$ms$) and 4.5 times faster than \textit{Normalization} (211$ms$ VS 47$ms$) for IPv6 verification}. The fundamental cause for such a large performance gap is due to the single trie traversal used in \textit{VeriTable} over a joint PT with intelligent selection of certain prefixes for comparisons without tree normalization, see Section~\ref{sec:verification} in detail. Note, that the leaf pushing operation over IPv6 forwarding table causes a significant inflation of the BTs which explains much slower speed  of \textit{TaCo} and \textit{Normalization} verification for IPv6 tables than for IPv4 tables.



\subsection{Number of Node Accesses}
\label{sec:accesses}

\begin{figure}[t]
		\captionsetup{justification=centering, width=0.49\linewidth}
		\subfloat[IPv4 Node Accesses]
		\centering
		\includegraphics[width=0.49\linewidth]{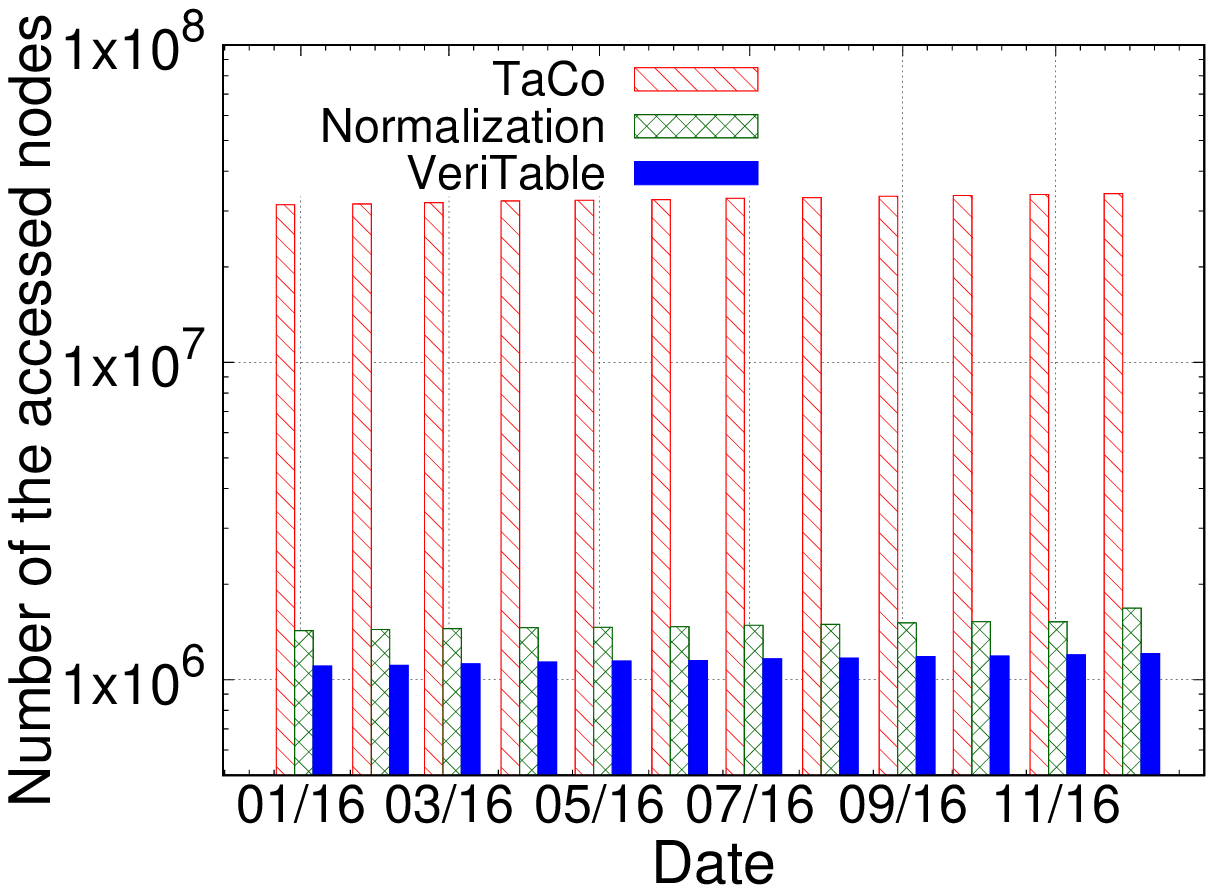}
		\label{fig:ipv4nodeaccesses}			
		\subfloat[IPv6 Node Accesses]
		\centering
		\includegraphics[width=0.49\linewidth]{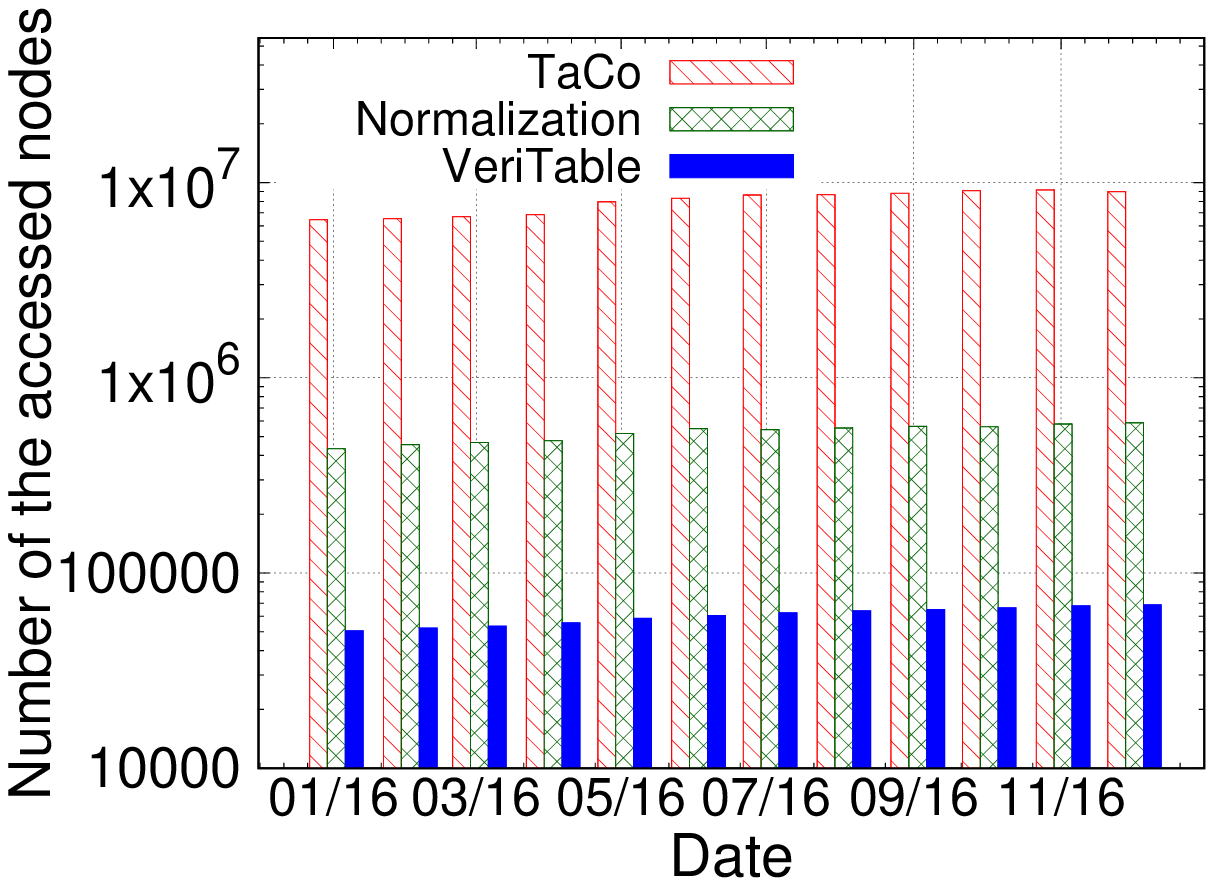}
		\label{fig:ipv6nodeaccesses}
	\vspace{-3mm}	
	\captionsetup{justification=centering, width=\linewidth}
	\caption{IPv4 and IPv6 Number of Node Accesses}
	\label{fig:accesses}
	\vspace{-3mm}
\end{figure}

Node accesses, similarly to memory accesses, refer to how many tree/trie nodes will be visited during verification. The total number of node accesses is the primary factor to determine the verification time of an algorithm. Figure~\ref{fig:accesses} shows the number of total node accesses for both IPv4 and IPv6 scenarios. Due to the novel design of \textit{VeriTable}, we are able to control the total number of node accesses to a significantly low level. For example, node accesses range from 1.1 to 1.2 million for 580K and 630K comparisons, which is \textbf{less than 2 node accesses} per comparison for IPv4, and it yields similar results for IPv6. On the contrary, \textit{TaCo} and \textit{Normalization} requires larger number of node accesses per comparison. For instance, \textit{TaCo} bears \textbf{35 node accesses} per comparison, on average, for IPv4 and \textbf{47 node accesses} per comparison, on average, in IPv6. \textit{Normalization} has \textbf{4 node accesses} per comparison in both cases. There are two main reasons for the gaps between \textit{VeriTable} and \textit{TaCo} and \textit{Normalization}: ($a$) \textit{VeriTable} uses a joint PT but \textit{TaCo} and \textit{Normalization} uses separate BTs. In a BT, it only goes one level down for each search step while multiple levels down in a PT; and ($b$) \textit{VeriTable} conducts only one post-order PT traversal. \textit{TaCo} conducts many repeated node accesses over a BT, including searching for a match from the BT root, using simply longest prefix matching process for each comparison. Due to the the unique form of a normalized BT, \textit{Normalization} requires no mutual IP address lookups and thus conducts significantly less node accesses than \textit{TaCo}.

\subsection{Memory Consumptions}
\label{sec:memoryconsumption}
\begin{figure}[t]
		\captionsetup{justification=centering, width=0.49\linewidth}
		\subfloat[IPv4 Memory Consumption]
		\centering
		\includegraphics[width=0.49\linewidth]{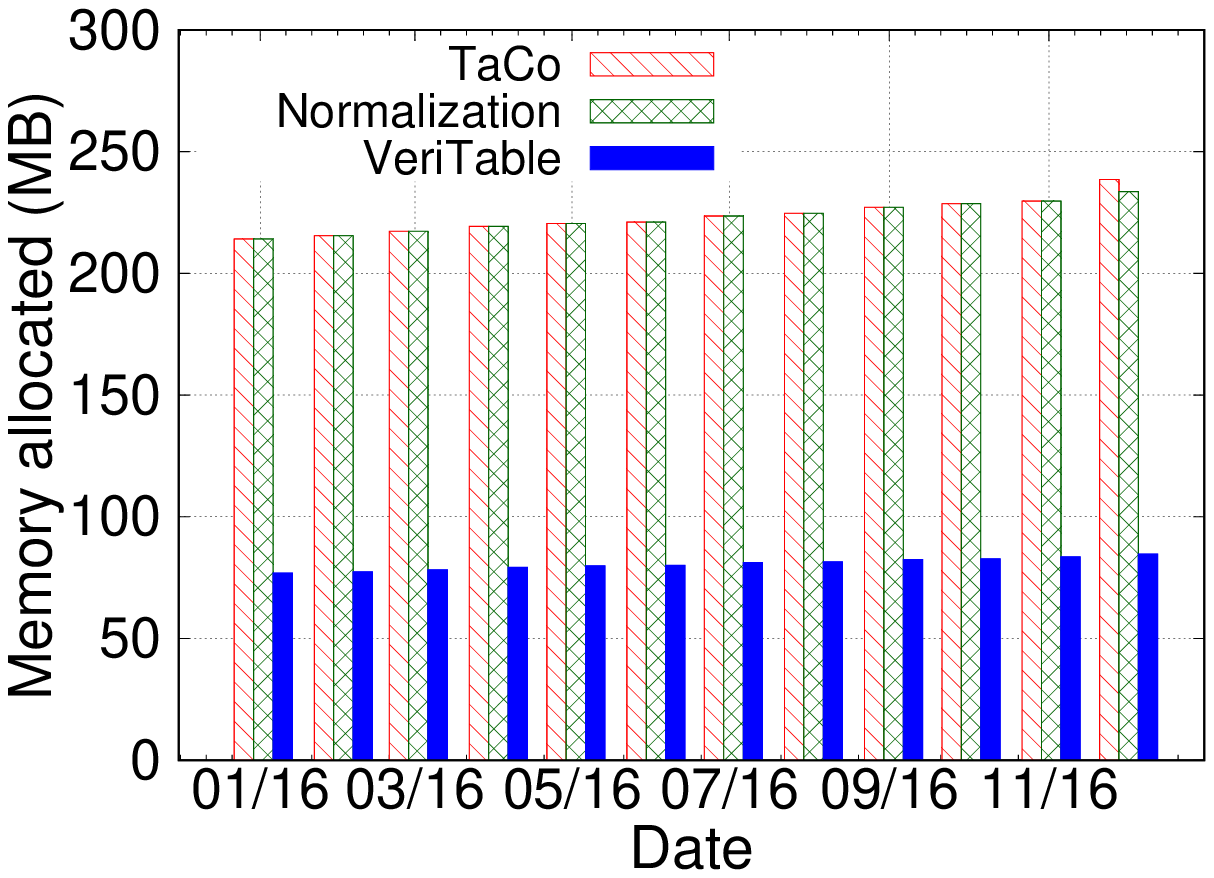}
		\label{fig:ipv4memory}	
		\subfloat[IPv6 Memory Consumption]
		\centering
		\includegraphics[width=0.49\linewidth]{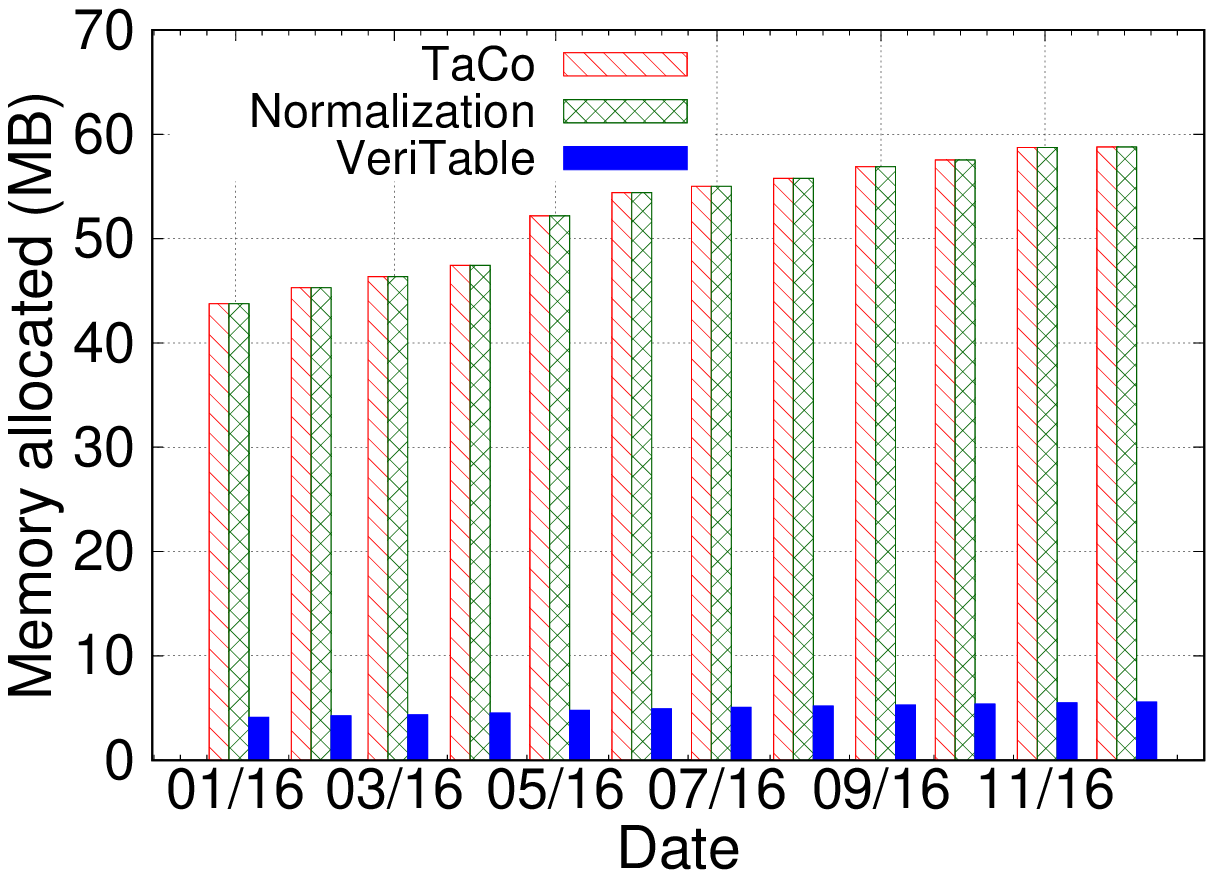}
		\label{fig:ipv6memory}
	\vspace{-3mm}
	\captionsetup{justification=centering, width=\linewidth}
	\caption{IPv4 and IPv6 Memory Consumption}
	\label{fig:memory}
	\vspace{-4mm}
\end{figure}

Memory consumption is another important metric to evaluate the performance of algorithms. Figure~\ref{fig:memory} shows the comparisons between \textit{TaCo}, \textit{Normalization} and \textit{VeriTable} for both IPv4 and IPv6 in terms of their total memory consumptions. 
In both scenarios, \textit{VeriTable} outperforms \textit{TaCo} and \textit{Normalization} significantly. \textbf{\textit{VeriTable} only consumes  around 38\% (80.86\textit{MB})} of total memory space than that of \textit{TaCo} and \textit{Normalization} (223\textit{MB}) on average for the same set of IPv4 forwarding tables. In the IPv6 case, \textit{VeriTable} bears even more outstanding results, \textbf{which only consumes 9.3\% (4.9\textit{MB})} of total memory space than that of \textit{TaCo} (53\textit{MB}) and \textit{Normalization} on average. Overall, thanks to the new design of our verification algorithm, \textit{VeriTable} outperforms \textit{TaCo} and \textit{Normalization} in all aspects, including total running time, number of node accesses and memory consumption. 

The differences in memory consumption by \textit{VeriTable}, \textit{Normalization} and \textit{TaCo} are caused by the unique combined trie data structure used in \textit{VeriTable}. A node in \textit{Normalization} and \textit{TaCo} holds a single next hop instead of an array of next hops, because \textit{TaCo} and \textit{Normalization} build separate BTs for each forwarding table. Moreover, those BTs inflate after leaf pushing.

\subsection{Performance for Multiple Tables}
\label{sec:multitable}

We also evaluated the performance of \textit{VeriTable} to check the forwarding equivalence and differences over multiple forwarding tables simultaneously. In the experiments, we intentionally added 2000 distinct errors when a new forwarding table was added. Then we verified that the same number of errors will be detected by \textit{VeriTable} algorithm. Starting from 2 tables, we gradually checked up to 10 tables simultaneously. The evaluation results have been shown in Table~\ref{tab:multitable}. There are two primary observations. First, \textit{VeriTable} is able to check the whole address space very quickly over 10 large forwarding tables (336.41$ms$) with relatively small memory consumptions (165$MB$). Second, the building time, verification time, node accesses, and memory consumptions grow much slower than the total number of forwarding entries. This indicates that \textit{VeriTable} can scale quite well for equivalence checking of a large number of tables. On the contrary, \textit{TaCo} and \textit{Normalization} naturally was not designed to compare multiple forwarding tables. In theory, \textit{TaCo} may need $n*(n-1)$ table-to-table comparisons to find the exact entries that cause differences, which is equal to 90 comparisons  for this 10-table scenario. On the other hand, \textit{Normalization} needs additional decompression steps to find such entries. We skip evaluation of \textit{TaCo} and \textit{Normalization} for multiple tables due to the high complexity. 

\subsection{``Blackholes" Detection}
\label{sec:blackhole}
A relaxed version with minor changes of our \textit{VeriTable} algorithm is able to quickly detect the routing space differences between multiple FIBs. More specifically, after building the joint PT for multiple FIBs, \textit{VeriTable} goes through the same verification process recursively. When traversing each \textit{Next Hops} array, it checks if there is a scenario where the array contains at least one default next hop (the next hop on default route 0/0) and at least one non-default next hop. If yes, it indicates that at least one FIB misses some routing space while another FIB covers it, which may lead to routing ``blackholes". In our experiments, we used data from RouteViews~\cite{routeviews} project, where 10 forwarding tables that contain the largest number of entries were collected and then merged to a super forwarding table with 691,998 entries. Subsequently, we one-to-one compared the routing spaces of the 10 individual forwarding tables with the super forwarding table. The results of these comparisons (see Table~\ref{tab:blackhole} in detail) show that none of these 10 forwarding tables fully cover the routing space of the merged one. The Leaking Routes in Table~\ref{tab:blackhole} were calculated by the number of subtrees in the joint PT under which an individual forwarding table ``leaks" certain routes but the merged super forwarding table covers them. These facts imply that the potential routing blackholes may take place between routers in the same domain or between different domains. To this end, our \textit{VeriTable} verification algorithm can identify these potential blackholes efficiently. On the contrary, \textit{TaCo} and \textit{Normalization} may not be easily used to detect block holes and loops  because different FIBs may result in different shapes of BTs (even normalized), which makes it hard for comparison. 



\begin{table*}[tbp]
	\centering
	\small
	\caption{Results of Comparing 10 IPv4 FIB Tables Simultaneously}
	\label{tab:multitable}
	\captionsetup{justification=centering}
	
	\begin{tabular}{| r | r | r | r | r | r | r | r |}
		\hline
		\thead{Number \\of tables} & \thead {Total number\\ of entries}&  \thead{Building\\ time($ms$)} & \thead{Verification \\time ($ms$)} & \thead{Number \\ of comparisons} & \thead{Node \\ access times} & \thead{Number \\ of errors}& Memory ($MB$) \\	
		\hline
		2 & 1230512 &962 & 82    &  586942 & 1133115 &4000& 84 \\ \hline
		3 & 1845768 &1326 & 108 &  1175884 & 1137115&6000& 94\\ \hline
		4 & 2461024 &1684 & 135 & 1766826 &  1141115 &8000& 104\\ \hline
		5 & 3076280 &2060 &  172 & 2359768 & 1145115&10000&114 \\  \hline
		6 & 3691536 &2471 & 194 &  2954710 & 1149115&12000& 124\\
		\hline
		7 & 4306792 &2869 & 213 &  3551652 & 1153115&14000& 134 \\ \hline
		8 & 4922048 &3248& 224 &  4150594 & 1157115&16000& 145\\ \hline
		9 & 5537304 &3630 & 322 &  4751536 & 1161115&18000& 155 \\  \hline
		10 & 6152560 &4007& 337 &  5354478 & 1165115&20000& 165\\
		\hline
	\end{tabular}
	\vspace{-5mm}
\end{table*}

%
%

\begin{table}[tbp]
	\centering
	\small
		\captionsetup{justification=centering}
	\caption{One-to-one Comparison of Individual \\Forwarding Tables with the Merged Super Table}
	\label{tab:blackhole}

	\begin{tabular}{| r | r | r |  r | r |}
		\hline
		Table size & Router IP & ASN  & \thead{BGP \\peers} & \thead{Leaking \\ Routes}  \\ \hline
		673083 & 203.189.128.233 &  23673  & 204 & 489\\ \hline
		667062 & 202.73.40.45 &	18106  & 1201 &  507\\  \hline
		658390 & 103.247.3.45 & 58511  & 1599 &  566\\ \hline
		657232 & 198.129.33.85 & 292  & 153 &  495 \\\hline
		655528 & 64.71.137.241 & 6939  &  6241 & 667 \\\hline
		655166 & 140.192.8.16 &  20130 & 2 &  879 \\\hline
		646912 &  85.114.0.217 & 8492  & 1504 &  796 \\ \hline
		646892 & 195.208.112.161 & 3277  & 4 &  772 \\ \hline
		641724 & 202.232.0.3 & 2497   & 294 &  1061 \\ \hline 
		641414 & 216.221.157.162 & 3257 & 316 &  1239 \\ \hline
		
	\end{tabular}
	\vspace{-2mm}
\end{table}

\

\section{Related Work}
\label{sec:related}
\textit{TaCo} algorithm, proposed by Tariq et al.~\cite{tariq2011taco}, is designed to verify forwarding equivalence between two forwarding tables. \textit{TaCo} builds two separate binary trees for two tables and performs tree normalization and leaf-pushing operations. Section~\ref{sec:stateart} elaborates the algorithm in detail. \textit{VeriTable} is very different from \textit{TaCo}. \textit{VeriTable} builds a single joint PATRICIA Trie for multiple tables and leverages novel ways to avoid duplicate tree traversals and node accesses, thus outperforms \textit{TaCo} in all aspects as shown in Section~\ref{sec:evaluation}. 
Inconsistency of forwarding tables within one network may lead to different types of problems, such as blackholes, looping of IP packets, packet losses and violations of forwarding policies. Network properties that must be preserved to avoid misconfiguration of a network can be defined as a set of invariants. Mai et al. introduced \textit{Anteater} in~\cite{mai2011debugging}, that converts the current network state and the set of invariants into instances of boolean satisfiability problem (SAT) and resolves them using heuristics-based SAT-solvers. Zeng et al. introduced \textit{Libra} in~\cite{zeng2014libra}, and they used MapReduce~\cite{dean2008mapreduce} to analyze rules from forwarding tables on a network in parallel. Due to the distributed model of MapReduce, \textit{Libra} analyzes the forwarding tables significantly faster than \textit{Anteater}. 
\textit{VeriFlow}~\cite{khurshid2012veriflow}, proposed by Khurshid et al., leverages software-defined networking to collect forwarding rules and then slice the network into \textit{Equivalence} \textit{classes} (\textit{EC}s). 
Kazemian et al. introduced \textit{NetPlumber} in \cite{kazemian2013real}, a real-time network analyzer based on \textit{Header Space Analysis }protocol-agnostic framework, described in \cite{kazemian2012header}. \textit{NetPlumber} is compatible with both SDN and conventional networks. It incrementally verifies the network configuration upon every policy change in a quick manner. 
Different from the network-wide verification methods above, \textit{VeriTable} aims to investigate whether multiple static forwarding tables yield the same forwarding behaviors-given any IP packet with a destination address or they cover the same routing space. 

\section{Conclusion}
\label{sec:conclusion}

We designed and developed \textit{VeriTable}, which can quickly determine if multiple routing or forwarding tables yield the same or different forwarding behaviors, and our evaluation results using real forwarding tables significantly outperform its counterparts. The novel algorithms and compact data structures can offer benefit not only in forwarding equivalence verification scenarios, but also in many other scenarios where we use Longest Prefix Matching rule for lookups, e.g., checking if route updates in control plane are consistent with the ones in forwarding plane. Moreover, the principles used in this paper can be applied to network-wide abnormality diagnosis of network problems, such as scalable and efficient forwarding loop detection and avoidance in the data plane of a network. In addition, VeriTable can be extended to handle incremental updates applied to the forwarding tables in a network. Our future work will explore these directions.

\bibliographystyle{IEEEtran}
\bibliography{net}
\end{document}